\documentclass[journal=jctcce,manuscript=article]{achemso}
\usepackage{amssymb,amsmath,amsfonts,multicol,multirow,longtable,array,mathpazo}
\usepackage{lscape}
\usepackage[version=3]{mhchem} 
\usepackage{appendix}
\usepackage{soul} 
\usepackage[usenames]{color}
\usepackage{makecell}
\usepackage{enumerate}
\usepackage{appendix}
\usepackage{xr}
\usepackage[T1]{fontenc} 
\usepackage{booktabs}
\usepackage{tikz}
\usepackage{threeparttable}
\usepackage{graphicx}
\usepackage{subfigure}
\usepackage{colortbl}
\usepackage{framed}
\usepackage{verbatim}
\usepackage{amsbsy}
\usepackage{bm}
\usepackage{bbm}
\usepackage[normalem]{ulem}
\usepackage{float}
\usepackage[linesnumbered,boxed]{algorithm2e}
\usepackage{algorithm2e}
\usepackage{amsmath}
\usepackage{diagbox}
\usepackage{changepage}

\mciteErrorOnUnknownfalse
\newcounter{xscheme}

\newcommand{\tr}{\mathrm{tr}}

\newfloat{Algorithm}{htbp}{alg}
\floatname{Algorithm}{Algorithm}

\newcounter{exe}[figure]
\newcommand{\iexe}{\refstepcounter{exe}\the\value{exe}:}

\setkeys{acs}{maxauthors = 0} 

\author{Yang Guo}
\affiliation{Qingdao Institute for Theoretical and Computational Sciences, Institute of Frontier and Interdisciplinary Science,
	Shandong University, Qingdao, Shandong 266237, China}
\author{Ning Zhang}
\affiliation{Beijing National Laboratory for Molecular Sciences, Institute of Theoretical and Computational Chemistry,
College of Chemistry and Molecular Engineering, Peking University, Beijing 100871, China}
\author{Wenjian Liu}\email{liuwj@sdu.edu.cn}
\affiliation{Qingdao Institute for Theoretical and Computational Sciences, Institute of Frontier and Interdisciplinary Science,
Shandong University, Qingdao, Shandong 266237, China}

\title{SOiCISCF: Combining SOiCI and iCISCF for Variational Treatment of Spin-orbit Coupling}
\setkeys{acs}{articletitle=true}

\begin{document}

\newpage

\begin{abstract}
It has recently been shown that the SOiCI approach [J. Phys.: Condens. Matter 34 (2022) 224007], in conjunction with the spin-separated exact two-component relativistic Hamiltonian,
can provide very accurate fine structures of systems containing heavy elements by treating electron correlation and spin-orbit coupling (SOC) on an equal footing.
Nonetheless, orbital relaxations/polarizations induced by SOC are not yet fully accounted for, due to the use of scalar relativistic orbitals.
This issue can be resolved by further optimizing the still real-valued orbitals self-consistently in the presence of SOC,
as done in the spin-orbit coupled CASSCF approach [J. Chem. Phys. 138 (2013) 104113]
but with the iCISCF algorithm [J. Chem. Theory Comput. 17 (2021) 7545] for large active spaces. The resulting SOiCISCF
employs both double group and time reversal symmetries for computational efficiency and assignment of target states.
The fine structures of $p$-block elements are taken as showcases to reveal the efficacy of SOiCISCF.

\end{abstract}

\maketitle

\clearpage
\newpage

\section{Introduction}

The simultaneous treatment of electron correlation and spin-orbit coupling (SOC) in electronic structure calculations
of atoms and molecules remains a challenge
to quantum chemistry\cite{liu2022perspective}. This is particularly true when a multi-configurational description of the system is mandatory.
The available multi-reference relativistic quantum chemical methods\cite{Fleig2012} can be classified into two big families, working with spinors or scalar orbitals.
The former try to treat SOC variationally already at the mean-field level, so as to account for SOC to all orders also in the subsequent treatment of correlation.
Such methods can further be classified into four-component (4C)\cite{4C-RASCI1993,4C-GASCI2003,4C-MCSCF1996,4C-CI-MCSCF2006,4C-MCSCF2008,
4C-CASPT22008,4C-CASSCF2015,4C-CASSCF2018,4C-icMRCI2015,4C-MR2018,4C-DMRG2014,4C-DMRG2018,
4C-DMRG2020,MPSSI2021,4c-FCIQMC,4C-MRMP21999,4C-CASPT22006} and two-component (2C)\cite{2C-CASSCF1996,2C-CI2001,2C-CASSCF2003,2C-CASSCF2013,LixiaosongX2CCASSCF,
2C-MRPT22014,LiXiaosongX2C-MRPT22022,2C-MRCI2020,X2C-DMRG-Li-2022} ones, which are computationally identical after integral transformations.
In this context, there exists even an intermediate approach, called quasi-four-component (Q4C)\cite{Q4C,X2C2007}, which looks like four-component
but computationally is precisely the same as a two-component approach even in the integral transformation step.
While such spinor-based (also called $jj$ coupling-based) methods are indispensable for core excitations and properties of systems containing very heavy elements,
they tend to `overshoot' valence excitations and properties, where SOC is usually not that strong,
such that it is the intermediate coupling that is more appropriate.
Here, $LS$-coupled many-electron basis functions of different spins are allowed to mix via SOC.
In other words, SOC is postponed to the correlation step, thereby permitting the use of real-valued
molecular orbitals (MO) and hence benefiting from nonrelativistic machinery and spin symmetry. For this reason, such approaches are usually called one-component (1C).
The interplay between SOC and electron correlation
can be accounted for in two ways\cite{Marian2001,marian2012SOCISC}, depending on what $LS$-coupled many-electron basis functions are used.
In the so-called one-step approach\cite{DGCIa,SpinUGA1,DGCIb,DGCIc,DetSOCI1997,CI-SOC1997,SPOCK2006,SOCASSCF2013,HBCISOC2017,CI-SOC1998,SOiCI,SlowTwoStep2022},
individual configuration state functions (CSF) of pre-chosen spins are used as the basis, so as to treat
spin-orbit and electron-electron interactions on an equal footing. In contrast, only a small number of low-lying correlated scalar states
are taken as the basis to construct an effective spin-orbit Hamiltonian matrix in the two-step approaches
\cite{CI-SOC1998,SOiCI,Hess1982,CIPSO1983,RASSOC,CASSI1986,CASSI1989,CI-SOC1997,SOCsingle-1,SOCsingle-2,SOCsingle-3,MRCI-SOC2000,
Teichteil2000,CASPT2-SOC2004,SI-SOC2006,EOMIPSO2008,mai2014perturbational,DMRGSO2015,
DMRGSO2016,nosiDMRGSO2016,Suo-SOC2021,vanWullen2021}.
That is, SOC is treated here after correlation. It is obvious that such two-step approaches
work well only when SOC and correlation are roughly additive. Of course, they become
identical with the one-step approaches if all correlated scalar states from the space
spanned by the individual CSFs are included. Since all the methods, whether 4C, 2C or 1C,
must converge to the same limit of full configuration interaction, it is clear that
the 1C approaches ought to handle a larger amount of spin-orbit-correlation energies
than the pure correlation energies in 4C/2C approaches, to compensate for the large portion of spin-orbit energies missed
in the mean-field step.
A direct deduction is then that orbit relaxations/polarizations induced by the spin-orbit interaction must be addressed properly
in the 1C approaches, so as to avoid the necessity of going to excitations of higher ranks than usually needed\cite{2C-CICC2012}.
Since the spin-orbit interaction is dominated by one-body terms,
it is clear that single excitations are most important for SOC. The questions here
are twofold: how to identify those `right' single excitations (SOSE for short) and how to determine their amplitudes.
For the former, both occupied and virtual orbitals are in question: It is often
the case that correlating more occupied orbitals (and hence more electrons) or more virtual orbitals actually worsens the SOC.
For atoms, it is well known\cite{CI-SOC1998,Teichteil2000} that the valence orbitals and
those virtual orbitals of the right shapes (i.e., with nodes close to the maximum of
the target scalar orbital and extrema on the two sides) are most important for
the biased spin polarization of the target scalar orbital towards the corresponding spinors. However, such a rule of thumb
is hardly useful for molecular systems. A more robust means is the virtual space decomposition (VSD) approach\cite{SDSRev,SOiCI},
which employs a medium-sized ECP (effective core potential) basis to map out
the right occupied and virtual orbitals from a large basis. It is particularly useful when a large, uncontracted all-electron basis is used,
where those virtual orbitals that are very high in energy but very local (HELO) in space must be screened out, especially when innermost core electrons are not to be correlated. As for the determination of the amplitudes of the SOSEs, it is well known\cite{2C-CICC2012,CCT1orbital1987} that only
the exponential of the SOSEs can do the job. However, multi-reference coupled-cluster theories are
very involved already in the absence of SOC, needless to say
in the presence of SOC. A much simpler way out is to optimize the
scalar orbitals self-consistently in the presence of SOC, as done in the spin-orbit coupled
CASSCF (complete active space self-consistent field) approach\cite{SOCASSCF2013}.
Following the same strategy, we here combine the iCISCF algorithm\cite{iCISCF} for large active spaces
with the recently proposed one-step 1C approach SOiCI\cite{SOiCI} to determine
the SOC-induced orbit relaxations/polarizations. Note in passing that
SOiCI arises from the combination of SOC with iCIPT2\cite{iCIPT2,iCIPT2New}, which itself is
a combination of the (exact) iterative configuration interaction (iCI)\cite{SDS,iCI,SDSRev} with configuration selection
for static correlation and second-order perturbation theory (PT2) for dynamic correlation.
The tabulated unitary group approach (TUGA)\cite{iCIPT2} is particularly effective
in the evaluation and reusage of basic coupling coefficients between randomly selected CSFs, even
for triplet excitation operators required by SOC\cite{SOiCI}. Both double group and time reversal symmetries can be employed
to facilitate the computation and assignment of the target states\cite{SOiCI}.

The rest of the paper is organized as follows. The SOiCISCF approach is described in Sec. \ref{Theory}. To make the presentation complete,
explicit expressions for the orbital gradient and Hessian are derived in Appendix \ref{AppGRDHESS}. Numerical results for prototypical systems
are then presented in Sec. \ref{Results} to reveal the efficacy of SOiCICSF. The paper is closed with concluding remarks in Sec. \ref{Conclusion}.
The following conventions are to be used throughout:
(1) the CI coefficients are generally complex but the orbitals are enforced to be real-valued;
(2) core, active, virtual, and arbitrary MOs are designated by $\{i,j,k,l,\cdots\}$, $\{e,f,t,u,v,w,\cdots\}$, $\{a,b,c,d,\cdots\}$, and $\{p,q,r,s,x,y,z\cdots\}$, respectively;
(3) repeated indices are always summed up.

\section{SOiCISCF}\label{Theory}
Given a set of real-valued MOs, the simplest variant of the spin-separated\cite{X2CSOCBook2017,X2CSOC1,X2CSOC2} exact two-component (X2C) Hamiltonians\cite{X2C2005,X2C2009}
can be written as
\begin{align}
\hat{H}&=\hat{H}^{\mathrm{sf}}+\hat{h}^{\mathrm{so}},\quad \hat{H}^{\mathrm{sf}}=\hat{h}^{\mathrm{sf}}+\hat{g}^{\mathrm{sf}},\label{Hoper}\\
\hat{h}^{\mathrm{sf}}&=h_{pq}^{\mathrm{sf}} \hat{E}_{pq},\quad \hat{E}_{pq}=a_{p\sigma}^{\dag}a_{q\sigma}=a^{p\sigma}_{q\sigma},\label{hsfOper}\\
\hat{g}^{\mathrm{sf}}&=\frac{1}{2} g_{pq,rs}\hat{e}_{pq,rs},\quad \hat{e}_{pq,rs}=a_{p\sigma}^\dag a_{r\tau}^\dag a_{s\tau} a_{q\sigma}=a^{p\sigma r\tau}_{q\sigma s\tau}=\hat{E}_{pq}\hat{E}_{rs}-\delta_{qr}\hat{E}_{ps},\\
 \hat{h}^{\mathrm{so}}&=\sum_{m=-1}^{1} Z_{pq}^{(-m)}\hat{T}^{(m)}_{pq}=\sum_{m=-1}^{1} Z_{pq}^{(m)}\hat{T}^{(-m)}_{pq},\\
 \mathbf{Z}^{(1)} & = i \mathbf{H}_{\mathrm{so}}^{x} - \mathbf{H}_{\mathrm{so}}^{y},\quad \mathbf{Z}^{(0)}= i \sqrt{2}\mathbf{H}_{\mathrm{so}}^{z}, \quad
\mathbf{Z}^{(-1)}= -i\mathbf{H}_{\mathrm{so}}^{x} - \mathbf{H}_{\mathrm{so}}^{y},\\
 \mathbf{Z}^{(m)T}&=-\mathbf{Z}^{(m)},\quad \mathbf{Z}^{(m)\dag}=-\mathbf{Z}^{(m)*}=(-1)^m \mathbf{Z}^{(-m)},\quad m=0,\pm 1,\\
\hat{T}_{pq}^{(1)} &= -a^{p\alpha}_{q\beta},\quad
\hat{T}_{pq}^{(0)} = \frac{1}{\sqrt{2}}(a^{p\alpha}_{q\alpha}-a^{p\beta}_{q\beta}), \quad
\hat{T}_{pq}^{(-1)} = a^{p\beta}_{q\alpha},\\
[\hat{T}^{(m)}_{pq}]^\dag&=(-1)^m \hat{T}^{(-m)}_{qp},\quad m=0,\pm 1,\\
\mathbf{H}_{\mathrm{so}}^l&=\mathbf{h}_{\mathrm{so,1e}}^l+\mathbf{f}_{\mathrm{so,2e}}^l, \quad(l=x,y,z),\label{HsoOper}
\end{align}
where $\mathbf{Z}^{(m)} $ and $\hat{T}_{pq}^{(m)}$ are rank-1 spherical tensor operators.
The explicit expressions for the real-valued, symmetric one-body spin-free (sf) integrals $h_{pq}^{\mathrm{sf}}$ in Eq. \eqref{hsfOper}
as well as the real-valued, antisymmetric, first-order effective one-body spin-orbit (so) integrals $(H_{\mathrm{so}}^{l})_{pq}$ ($l = x, y, z$)
in Eq. \eqref{HsoOper} are well documented in Ref. \citenum{SOiCI}
and hence not repeated here.

After partitioning the MOs into core, active, and virtual ones, the selection of individual CSFs of spin $S$ in the CASCI space $P$ is
performed iteratively until convergence with respect to static correlation (see equations (37) and (39) in Ref. \citenum{SOiCI} for the selection criteria). Additional singly excited CSFs of spins $S-1$, $S$, and $S+1$ that are important for SOC are then selected
in one step (see equation (38) in Ref. \citenum{SOiCI} for the selection criterion). It is possible to
select simultaneously those CSFs important for both static correlation and SOC but this is hardly needed.
A single parameter $C_{\mathrm{min}}$ is invoked to control the size of the final, selected space $P_m$. The Hamiltonian \eqref{Hoper} projected onto $P_m$
is finally diagonalized\cite{iVI,iVI-TDDFT} to yield the normalized SOiCI wave function
\begin{eqnarray}
|0\rangle =\sum_{|I\rangle \in P_m} |I\rangle C_I^{(0)},\quad \langle I|J\rangle=\delta_{IJ}, \quad \sum_I C_I^{(0)*}C_I^{(0)}=1 \label{iCI}
\end{eqnarray}
with energy
\begin{align}
E^{(0)}&=\tr(\gamma h^{\mathrm{sf}} ) +\frac{1}{2}\sum_{rs}\tr(\mathbf{P}^{rs}\mathbf{J}^{rs})-\sum_{m=-1}^{1}\tr (\mathbf{S}^{(m)}\mathbf{Z}^{(-m)}),\label{SymmE}\\
h^{\mathrm{sf}}_{pq}&=h^{sf}_{qp},\quad J^{rs}_{pq}=g_{pq,rs}=J^{sr}_{pq}=J^{rs}_{qp}=J^{sr}_{qp},\quad Z_{pq}^{(m)}=-Z_{qp}^{(m)},\\
D_{pq}&=\langle 0|\hat{E}_{pq}|0\rangle=D_{qp}^*,\quad \gamma_{pq}=\frac{1}{2}(D_{pq}+D_{qp})=\Re D_{pq},\\
S_{pq}^{(m)}&=\langle 0|\hat{T}^{(m)}_{pq}|0\rangle=(-1)^m S_{qp}^{(-m)*},\label{Smat} \\
\Gamma_{pq,rs}&=\langle 0|\hat{e}_{pq,rs}|0\rangle=\Gamma_{qp,sr}^*,\\
P^{rs}_{pq}&=\frac{1}{4}(\Gamma_{pq,rs}+\Gamma_{qp,rs}+\Gamma_{pq,sr}+\Gamma_{qp,sr}) \\
&=\frac{1}{2}\Re(\Gamma_{pq,rs}+\Gamma_{qp,rs})=P^{rs}_{qp}=P^{sr}_{pq}=P^{sr}_{qp}.
\end{align}
Eq. \eqref{SymmE} (see also Eq. \eqref{FinalE}) arises because the orbitals are enforced to be real-valued and spin adapted (i.e., same orbital for different spins).
However, the CI coefficients are in general complex due to the presence of the SOC operator $\hat{h}^{so}$.
Note in passing that the basic coupling coefficients $\langle I |\hat{E}_{pq}|J\rangle$, $\langle I |\hat{T}^{(m)}_{pq}|J\rangle$, and $\langle I|\hat{e}_{pq,rs}|J\rangle$ in
the Hamiltonian matrix elements $\langle I|\hat{H}|J\rangle$ depend only on the relative occupations of the MOs in the CSF pairs $\{|I\rangle, |J\rangle\}$ but not on
the individual MOs. As such, they can be evaluated and reused very efficiently with TUGA\cite{iCIPT2,SOiCI}.
In particular, both double group and time reversal symmetries have been incorporated\cite{SOiCI} into the evaluation of $\langle I|\hat{H}|J\rangle$.
For later use, we here define $P_s=P-P_m$ as the residual of $P$ left over by the selection, whereas $Q=1-P$ as the complementary space for dynamic correlation.

To further optimize the orbitals but retaining both point group and spin symmetries (which
are necessary for subsequent treatment of dynamic correlation within the 1C framework), we parameterize the SOiCISCF wave function as\cite{ElectronStructure}
\begin{align}
|\tilde{0}\rangle &=\sum_{I\in P_m}e^{-\hat{\kappa}}|I\rangle C_I,\label{RotPara2} \\
\hat{\kappa}&=\sum_{pq}\kappa_{pq}\hat{E}_{pq}=\sum_{p>q}\kappa_{pq} \hat{E}_{pq}^-=\frac{1}{2}\sum_{p,q}\kappa_{pq}\hat{E}_{pq}^-,\label{kappadef}\\
\boldsymbol{\kappa}&=-\boldsymbol{\kappa}^\dag,\quad
\hat{E}_{pq}^- = \hat{E}_{pq}-\hat{E}_{qp}.\label{RotOp}
\end{align}
The skew symmetry of $\boldsymbol{\kappa}$ implies that only (real-valued) $\kappa_{pq} (p>q)$, with $p$ and $q$
belonging to the same irreducible representation (irrep) of binary point groups ($D_{2h}$ and its subgroups),
 are independent parameters. Moreover, rotations within the core and within the virtual orbitals are also redundant. Therefore, the $\hat{\kappa}$ operator
actually reads
\begin{eqnarray}
\hat{\kappa}=
\sum_{ai} \kappa_{ai}\hat{E}_{ai}^- + \sum_{ti} \kappa_{ti}\hat{E}_{ti}^- + \sum_{at} \kappa_{ta}\hat{E}_{at}^-+ \sum_{t>u} \kappa_{tu}\hat{{E}}_{tu}^-,\label{redKappa}
\end{eqnarray}
where the last term vanishes in CASSCF but has to be included in SOiCISCF due to the truncation of CASCI.
To optimize the CI coefficients and orbitals, we construct the following Lagrangian
\begin{eqnarray}
L[\mathbf{C},\mathbf{C}^*,\boldsymbol{\kappa}]&=&\langle \tilde{0}|\hat{H}|\tilde{0}\rangle -E(\mathbf{C}^\dag \mathbf{C}-\mathbf{I}). \label{Lagrangian}
\end{eqnarray}
Since $L$ \eqref{Lagrangian} is quadratic in the CI coefficients, the latter can be obtained by diagonalizing the CI matrix $\mathbf{H}$ for a given set of MOs.
As a result, the stationarity conditions
\begin{align}
G^{\mathrm{c*}}_I &=\frac{\partial L}{\partial C_I^*}\vline_{\boldsymbol{\xi}=\boldsymbol{\xi}^{(0)}}=\langle I|\hat{H}-E^{(0)}|0\rangle=(G^{\mathrm{c}}_I)^*=0\label{CICond}
\end{align}
for the CI coefficients are satisfied automatically for fixed orbitals.
To further determine changes in the orbitals with given CI coefficients, we expand
the Lagrangian $L$ to second order at the expansion point $\boldsymbol{\xi}^{(0)}=(\mathbf{C}^{(0)}, \mathbf{C}^{(0)*},\mathbf{0})^T$,
\begin{align}
L^{(2)}[\mathbf{C},\mathbf{C}^*,\boldsymbol{\kappa}]&=E^{(0)}+\boldsymbol{\xi}^{(1)\dag} \mathbf{G}+\frac{1}{2}\boldsymbol{\xi}^{(1)\dag} \mathbf{E} \boldsymbol{\xi}^{(1)},
\end{align}
the stationary condition of which is
\begin{eqnarray}
\mathbf{E}\boldsymbol{\xi}^{(1)}=-\mathbf{G},
\end{eqnarray}
or in block form
\begin{eqnarray}
\begin{pmatrix}
\mathbf{E}^{\mathrm{c^*c}}&\mathbf{E}^{\mathrm{c^*c^*}}&\mathbf{E}^{\mathrm{c^*o}}\\
\mathbf{E}^{\mathrm{cc}}&\mathbf{E}^{\mathrm{cc^*}}&\mathbf{E}^{\mathrm{co}}\\
\mathbf{E}^{\mathrm{oc}}&\mathbf{E}^{\mathrm{oc^*}}&\mathbf{E}^{\mathrm{oo}}\end{pmatrix}
\begin{pmatrix}\mathbf{C}^{(1)}\\ \mathbf{C}^{(1)^*} \\ \boldsymbol{\kappa}^{(1)}
\end{pmatrix}=-\begin{pmatrix} \mathbf{G}^{\mathrm{c^*}}\\ \mathbf{G}^{\mathrm{c}}\\ \mathbf{G}^{\mathrm{o}}\end{pmatrix},
\quad \mathbf{G}^{\mathrm{c}}=\mathbf{G}^{\mathrm{c^*}}=\mathbf{0},\label{NReq}
\end{eqnarray}
where the Hessian and orbital gradient read
\begin{align}
E^{\mathrm{c^*c}}_{IJ}&= \frac{\partial^2 L}{\partial C_I^* \partial C_J}\vline_{\boldsymbol{\xi}=\boldsymbol{\xi}^{(0)}}=\langle I|\hat{H}-E^{(0)}|J\rangle=(E^{\mathrm{cc*}}_{IJ})^*,\label{Ecc}\\
E^{\mathrm{c^*c^*}}_{IJ}&= \frac{\partial^2 L}{\partial C_I^* \partial C_J^*}\vline_{\boldsymbol{\xi}=\boldsymbol{\xi}^{(0)}}=0=E^{\mathrm{cc}}_{IJ},\\
E^{\mathrm{c^*o}}_{I,rs}&=\frac{\partial^2 L}{\partial  C_I^* \partial \kappa_{rs}}\vline_{\boldsymbol{\xi}=\boldsymbol{\xi}^{(0)}}=\langle I|[\hat{E}_{rs}^-, \hat{H}]|0\rangle=(E^{\mathrm{co}}_{I,rs})^*, \label{Eco}\\
E^{\mathrm{oc^*}}_{pq,I}&=\frac{\partial^2 L}{\partial \kappa_{pq} \partial C_I^*}\vline_{\boldsymbol{\xi}=\boldsymbol{\xi}^{(0)}}=\langle I|[\hat{E}_{pq}^-, \hat{H}]|0\rangle=(E^{\mathrm{oc}}_{pq,I})^*,\label{Eoc}\\
E^{\mathrm{oo}}_{pq,rs}&=\frac{\partial^2 L}{\partial \kappa_{pq} \partial \kappa_{rs}}\vline_{\boldsymbol{\xi}=\boldsymbol{\xi}^{(0)}}
=\frac{1}{2}\langle 0|[\hat{E}_{pq}^-,[\hat{E}_{rs}^-,\hat{H}]]|0\rangle+\frac{1}{2}\langle 0|[\hat{E}_{rs}^-,[\hat{E}_{pq}^-,\hat{H}]]|0\rangle,\label{EOO1}\\
&=\langle 0|[\hat{E}_{pq}^-,[\hat{E}_{rs}^-,\hat{H}]]|0\rangle+\frac{1}{2}\langle 0|[[\hat{E}_{rs}^-,\hat{E}_{pq}^-],\hat{H}]|0\rangle=E^{\mathrm{oo}*}_{pq,rs}\\
&=-E^{\mathrm{oo}}_{qp,rs}=-E^{\mathrm{oo}}_{pq,sr}=E^{\mathrm{oo}}_{qp,sr}=E^{\mathrm{oo,sf}}_{pq,rs}+E^{\mathrm{oo,so}}_{pq,rs},\label{Orbhessian}\\
G^{\mathrm{o}}_{pq}&=\frac{\partial L}{\partial \kappa_{pq}}\vline_{\boldsymbol{\xi}=\boldsymbol{\xi}^{(0)}}=\langle 0|[\hat{E}_{pq}^{-}, \hat{H}]|0\rangle=G^{\mathrm{o}*}_{pq}=-G^{\mathrm{o}}_{qp}=G^{\mathrm{o,sf}}_{pq}+G^{\mathrm{o,so}}_{pq}.\label{Orbgrd}
\end{align}
The explicit expressions for the orbital gradient $G_{pq}^{\mathrm{o}}$ and Hessian $E^{\mathrm{oo}}_{pq,rs}$ are documented in Appendix \ref{AppGRDHESS}.
The first (and second) row of Eq. \eqref{NReq} can be rearranged to
\begin{eqnarray}
(\mathbf{H}^{(0)}-E^{(0)}\mathbf{I})\mathbf{C}^{(1)}=-\mathbf{G}^{\mathrm{c^*}}-\mathbf{H}^{(1)}\mathbf{C}^{(0)},\label{CIresp}
\end{eqnarray}
where
\begin{align}
H^{(1)}_{IJ}&=\langle I|\hat{H}_{\kappa}|J\rangle,\label{H1mat}\\
\hat{H}_{\kappa}&=[\hat{\kappa},H]=\frac{1}{2}\kappa_{pq}^{(1)}[\hat{E}_{pq}^-,\hat{H}]=\hat{H}^{\dag}_{\kappa},\quad \forall p,q\label{Hkoper}\\
&=[\boldsymbol{\kappa}^{(1)},\mathbf{h}^{\mathrm{sf}}]_{pq}\hat{E}_{pq}+[\boldsymbol{\kappa}^{(1)},\mathbf{J}^{rs}]_{pq}\hat{e}_{pq,rs}+\sum_{m=-1}^{1} [\boldsymbol{\kappa}^{(1)},\mathrm{Z}^{(-m)}]_{pq}\hat{T}^{(m)}_{pq}.\label{Hkappa}
\end{align}
Use of Eqs. \eqref{Epq-Hsf} and \eqref{Epq-Hso} has been made when going from Eq. \eqref{Hkoper} to Eq. \eqref{Hkappa}.
Likewise, the third row of Eq. \eqref{NReq} can be rearranged to
\begin{align}
(\mathbf{E}^{\mathrm{oo}}\boldsymbol{\kappa}^{(1)})_{pq}&=-\mathbf{G}^{\mathrm{o}}_{pq}-2\Re(\langle 0|[\hat{E}_{pq}^-,\hat{H}]|J\rangle C_J^{(1)}),\label{Orbresp}\\
(\mathbf{E}^{\mathrm{oo}}\boldsymbol{\kappa}^{(1)})_{pq}&= \langle 0|[\hat{E}_{pq}^-,\hat{H}_{\kappa}]|0\rangle+[\mathbf{G^{\mathrm{o}}},\boldsymbol{\kappa}^{(1)}]_{pq},\label{OrbOrbsigma}\\
\langle 0|[\hat{E}_{pq}^-,\hat{H}|J\rangle C_J^{(1)}&=2[\boldsymbol{\gamma}^{(1)},\mathbf{h}^{\mathrm{sf}}]_{pq}+2\sum_{rs}[\mathbf{P}^{(1)rs},\mathbf{J}^{rs}]_{pq}
-2\sum_{m=-1}^{1}[\mathbf{t}^{(1,m)},\mathbf{Z}^{(-m)}]_{pq},\label{OrbCIsigma}\\
\gamma^{(1)}_{pq}&=\frac{1}{2}\langle 0|\hat{E}_{pq}+\hat{E}_{qp}|J\rangle C_J^{(1)},\label{Den1}\\
P^{(1)rs}_{pq}&=\frac{1}{2} \langle 0|\hat{e}_{pq,rs}+\hat{e}_{qp,rs}|J\rangle C_J^{(1)},\label{Gamma1}\\
t^{(1,m)}_{pq}&=\frac{1}{2}(1-P_{pq})\langle 0|\hat{T}^{(m)}_{pq}|J\rangle C_J^{(1)}.
\end{align}
Eqs. \eqref{CIresp} and \eqref{Orbresp} can be viewed as response equations for the CI displacements $\mathbf{C}^{(1)}$ and orbital Newton steps $\boldsymbol{\kappa}^{(1)}$, respectively.
The former involves the first-order active space Hamiltonian \eqref{H1mat} due to $\boldsymbol{\kappa}^{(1)}$, whereas the latter involves the first-order reduced density matrices (RDM) \eqref{Den1}/\eqref{Gamma1} due to $\mathbf{C}^{(1)}$.
This particular reformulation\cite{DMRGSCF2017b} is advantageous in that it decouples the orbital optimization from the CI solver implementation in each Newton step, such that any CI solver can readily be used.
Considering that the simultaneous optimization of the CI coefficients and the orbitals suffers
from severe linear dependence due to the presence of active-active orbital rotations and that the CI step is rate determining for a large CAS,
 the second term on
the right-hand side of both Eqs. \eqref{CIresp} and \eqref{Orbresp} can be ignored, thereby leading to
\begin{align}
(\mathbf{H}^{(0)}-E^{(0)}\mathbf{I})\mathbf{C}^{(1)}&=-\mathbf{G}^{\mathrm{c^*}},\label{CIresp0}\\
\mathbf{E}^{\mathrm{oo}}\boldsymbol{\kappa}^{(1)}&=-\mathbf{G}^{\mathrm{o}}. \label{Orbresp0}
\end{align}
Eq. \eqref{CIresp0} is the usual iterative partial diagonalization of the CI matrix $\mathbf{H}^{(0)}$, keeping the orbitals fixed. For the given RDMs,
Eq. \eqref{Orbresp0} can, in the spirit of quasi-Newton (QN) methods, be solved iteratively
with the Broyden-Fletcher-Goldfarb-Shanno (BFGS) algorithm\cite{BFGSorg,BFGS} (microiteration), which amounts to updating iteratively the inverse of the orbital Hessian $\mathbf{E}^{\mathrm{oo}}$ with
the diagonal elements (see Appendix \ref{AppGRDHESS}) as initial guesses. However, such a QN scheme
is effective only for the optimization of rotations between core/active and virtual and between core and active orbitals.
Instead, the Jacobi rotation (JR) algorithm\cite{Jacobi} is more robust for the optimization of active orbitals.
As such, it is the QR+JR hybrid that is preferred\cite{iCISCF}.
The spin-orbit interaction may be switched on only after sf-X2C-iCISCF has converged.
If a state-specific Epstein-Nesbet type of second-order perturbation theory (ENPT2) is carried out in $P_s$, the method will be denoted as SOiCISCF(2),
which can be made very close to SOCASSCF by reducing the cutoff thresholds
($C_{\mathrm{min}}$ for configuration selection and $A_{\mathbf{min}}$ for active orbital rotations).

\section{Computational details}
The SOiCISCF approach has been implemented in a development version of the BDF program package\cite{BDF1,BDF2,BDF3,BDFrev2020}.
For all iCISCF and SOiCISCF calculations with 14 active orbitals or less,
the threshold $C_{\mathrm{min}}$ for configuration selection
is set to zero (i.e., iCISCF=CASSCF and SOiCISCF=SOCASSCF), whereas
for calculations with 21 active MOs or less, $C_{\mathrm{min}}$ is set to 10$^{-7}$. For calculations with even more active MOs,
$C_{\mathrm{min}}$ is set to 10$^{-5}$. However, extrapolated results will also be provided when all virtual MOs are included
in the active space.
A value of 10$^{-6}$ is always employed for $A_{\mathrm{min}}$.
Although only $D_{2h}$ symmetry is employed in the atomic calculations,
full orbital degeneracy is guaranteed by allowing only averaged rotations between orbitals of the same angular momentum, i.e.,
$\kappa_{nl_m,n^\prime l_m}=\frac{1}{2l+1}\sum_{i=-l}^l \kappa_{nl_i,n^\prime l_i}$, $m\in[-l,l]$.
The [Ne], [Ar], and [Kr] cores are kept frozen for Br, I, and At, respectively.
Other $p$-block elements are treated similarly.

\section{Results and Discussion}\label{Results}

\subsection{Basis set effects} \label{ssec basis}
The Dyall triple-zeta basis sets\cite{dyalltz} are recontracted here as follows. The radial sf-X2C-HF equation is first solved
for each spherically averaged, non-polarized atomic configuration. The coefficients of the so-obtained radial atomic orbitals (AO) are
then used to generate generally contracted basis functions for the occupied AOs. The first and third $s$ and $p$,
the first and second $d$, as well as the first $f$ primitives (counted from the smallest to
the largest exponents) are further added,
leading to `HF+P' type of polarized valence triple-zeta (vtz) scalar basis sets. Such basis sets are known
to be good enough for valence properties in the absence of SOC. However, they have sizable errors for spin-orbit splittings (especially of $np$ orbitals),
as compared with the parent uncontracted basis sets\cite{X2CSOC1}. The underlying reason is that the $np_{1/2}$ spinors are radially very
different from the scalar $np$ orbitals, such that they cannot be well described by truncated,
contracted scalar basis functions.
The situation is much less severe for other types of spinors. One way out is to further add in some steep functions\cite{van1999geometry,nicklass2000convergence,wang2001error,armbruster2006basis}. However, this requires
nonlinear optimizations for each basis set. A more effective means\cite{X2CSOC1} is to
take the $np$ contracted functions that are directly
projected out from the $\alpha$ components of $np_{1/2}$ as the additional functions.
The so-augmented basis sets are to be denoted as vtz-$m$p$^1_2$, with $m$ being the number of additional scalar
functions (e.g., $m=3$ if the additional scalar functions are projected out from $2p_{1/2}$, $3p_{1/2}$, and $4p_{1/2}$).
Similar functions projected out from other spinors (e.g., the lowest $2p_{3/2}$ and $3d_{3/2}$ spinors) are hardly needed
and hence not to be considered here.

\begin{figure}[!htp]
	\footnotesize
	\centering
	\begin{tabular}{cc}
		\includegraphics[width=0.4\textwidth]{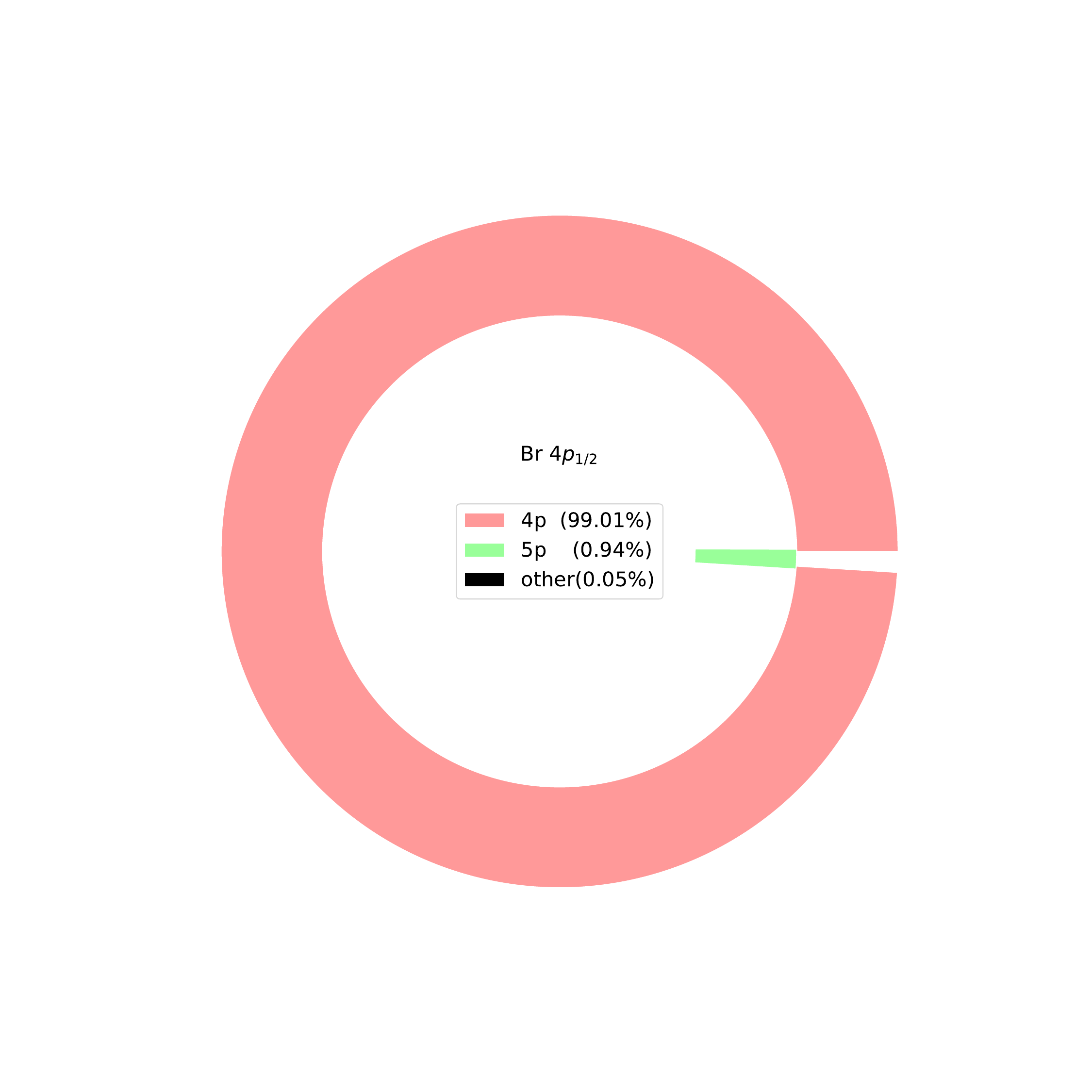}&\includegraphics[width=0.4\textwidth]{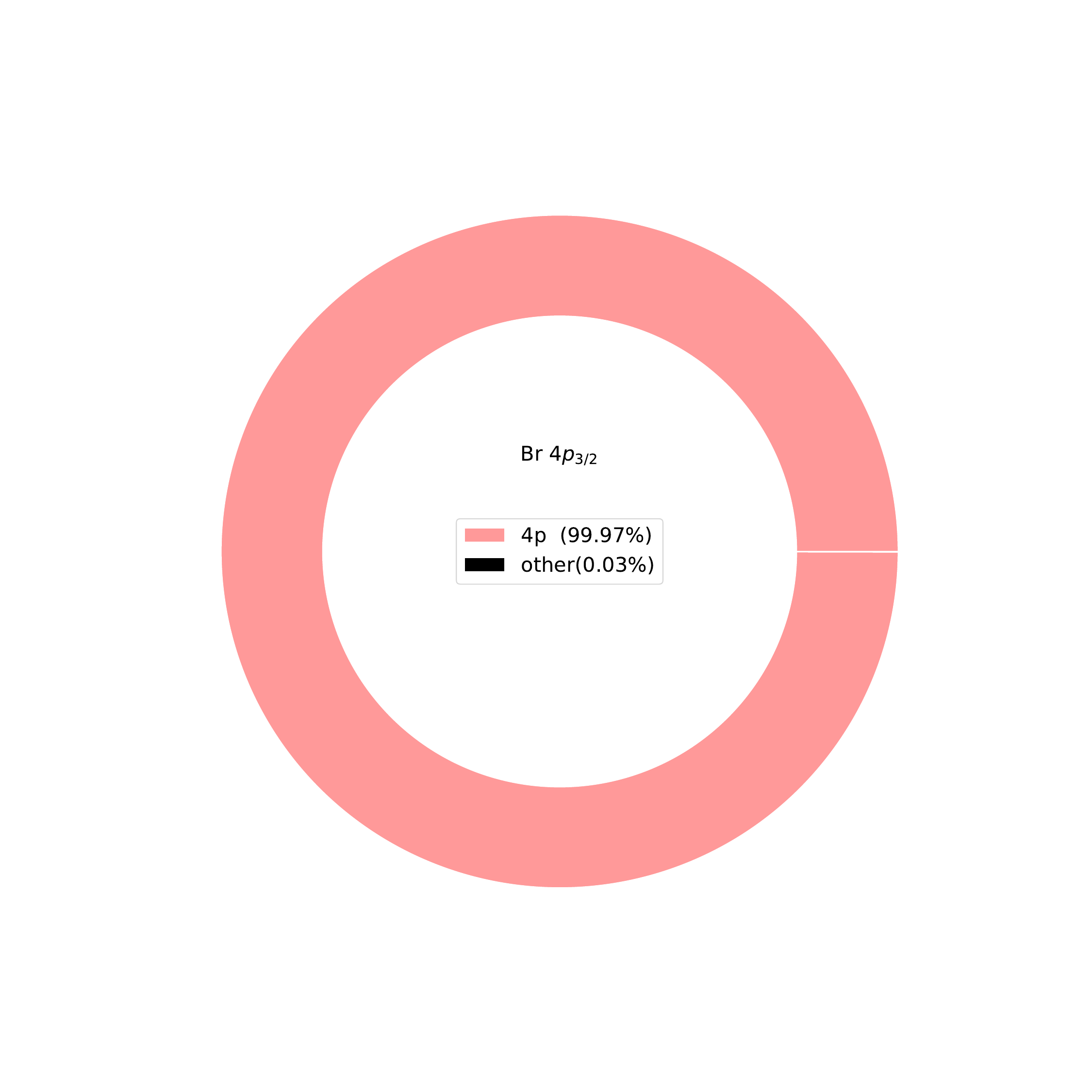}\\
		\includegraphics[width=0.4\textwidth]{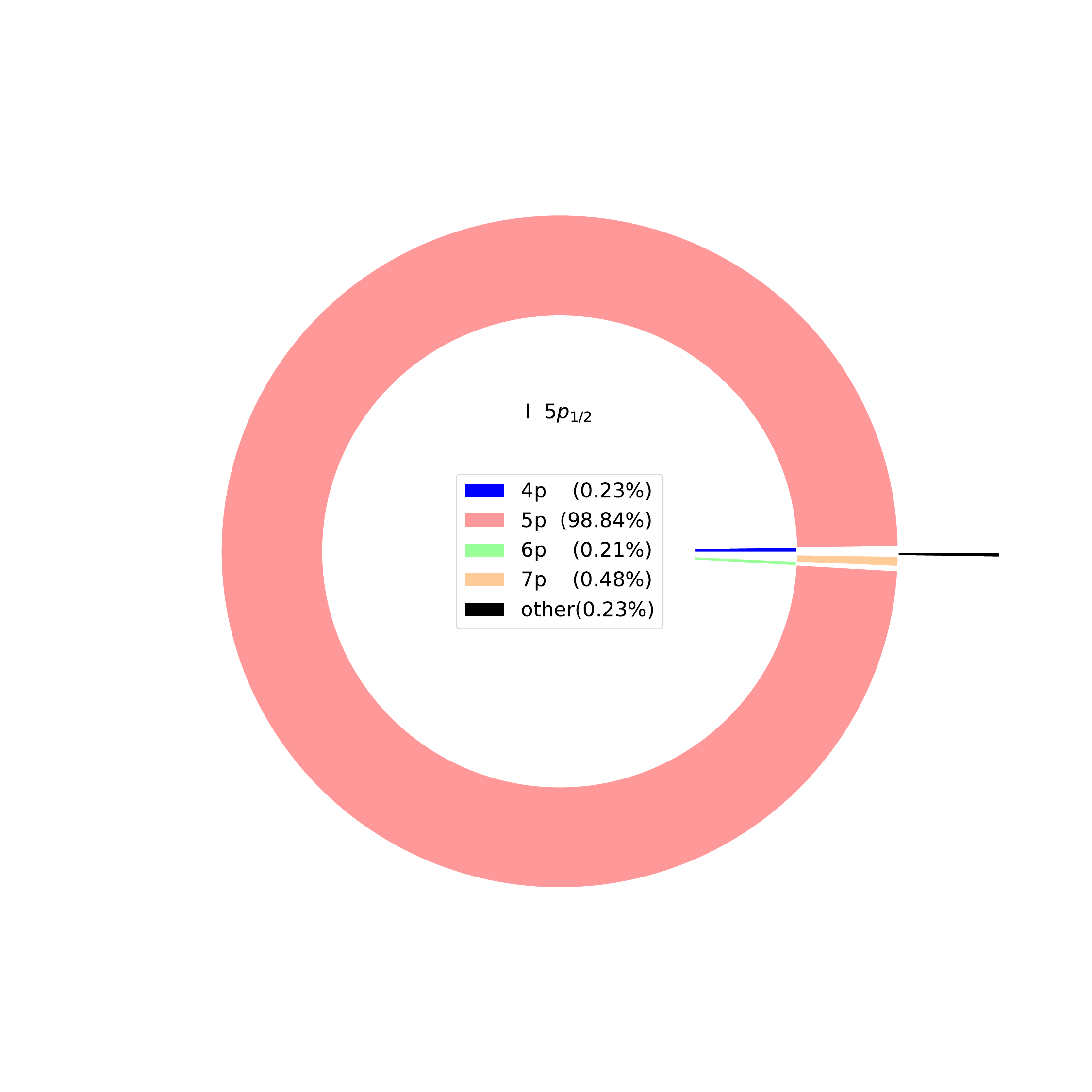}&\includegraphics[width=0.4\textwidth]{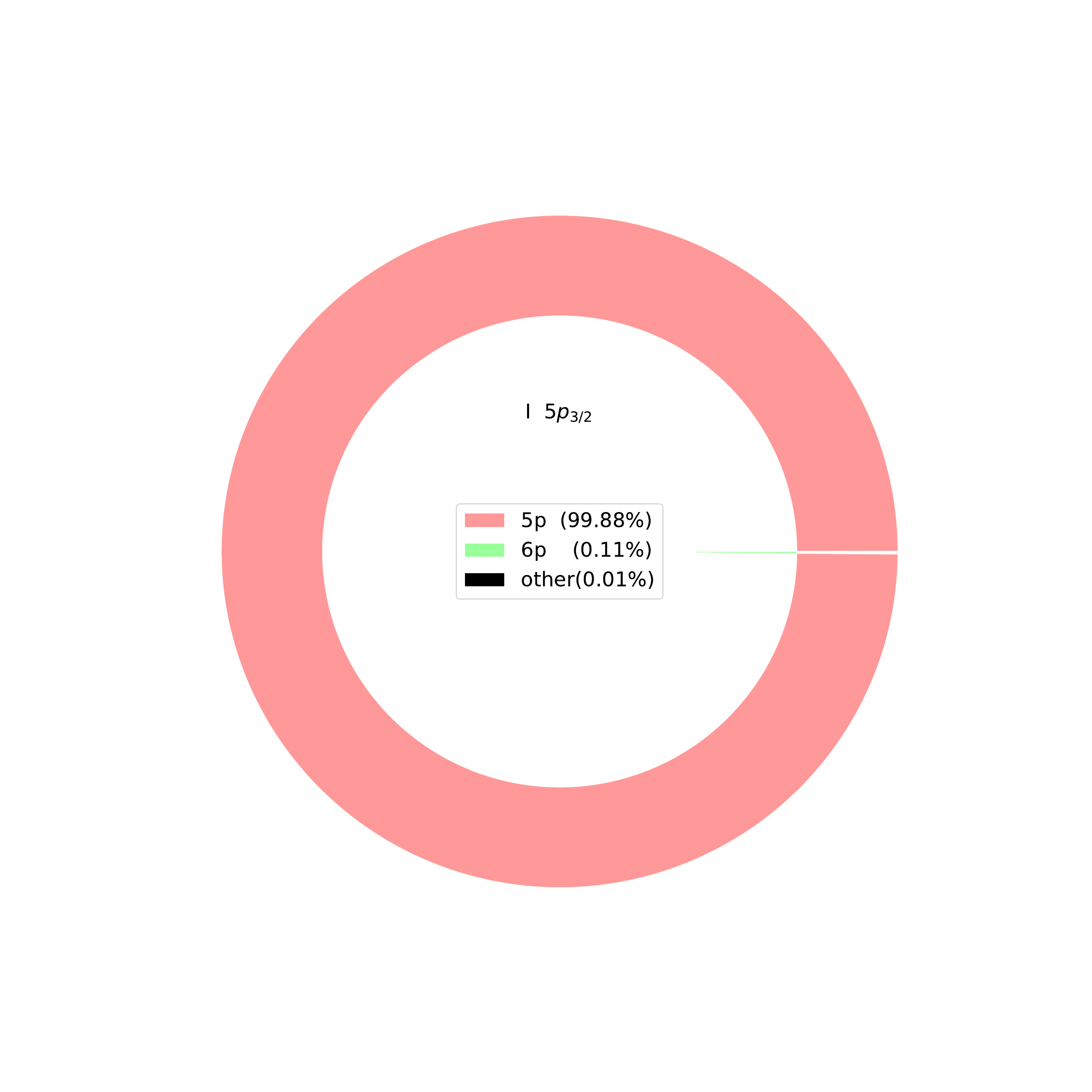}\\
		\includegraphics[width=0.4\textwidth]{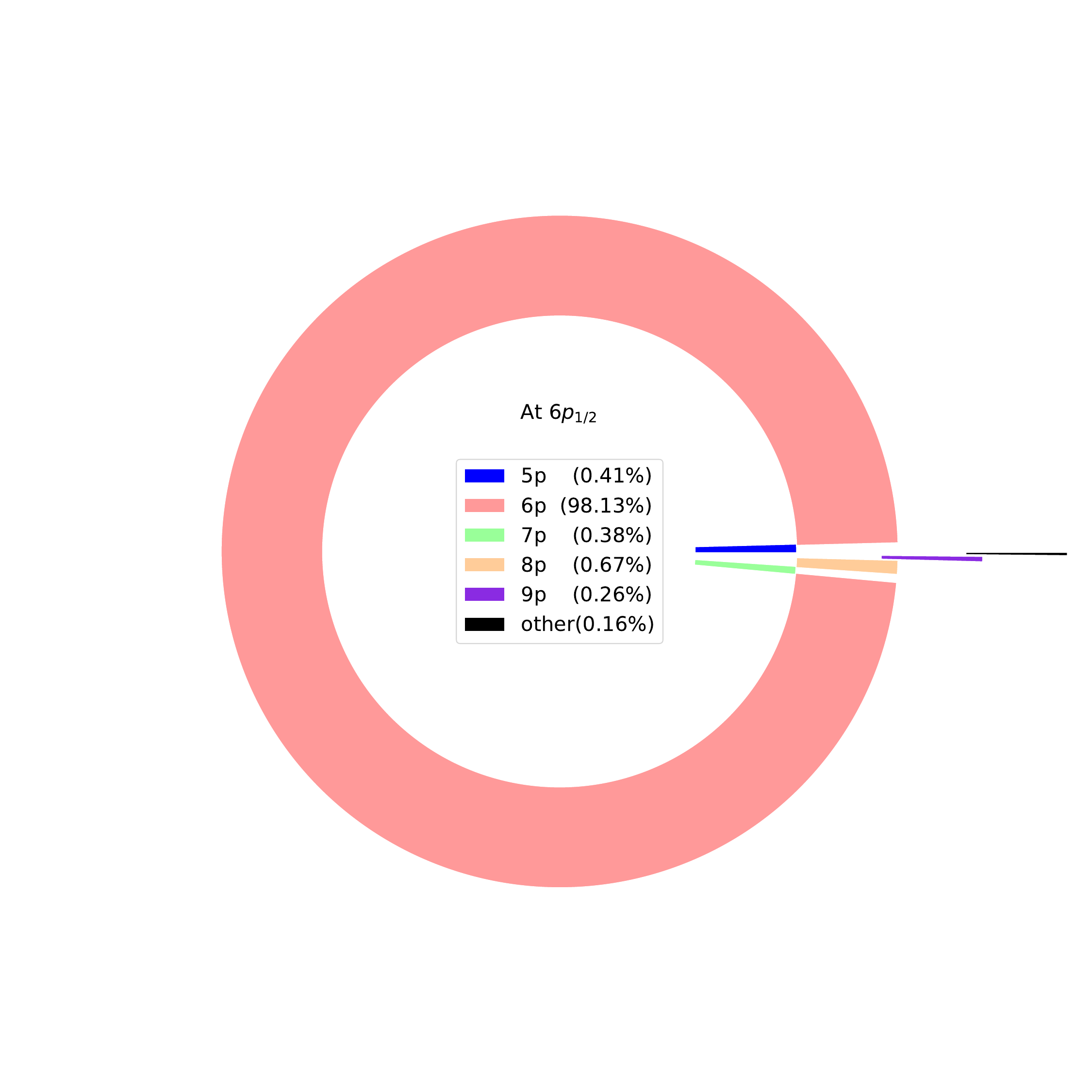}&\includegraphics[width=0.4\textwidth]{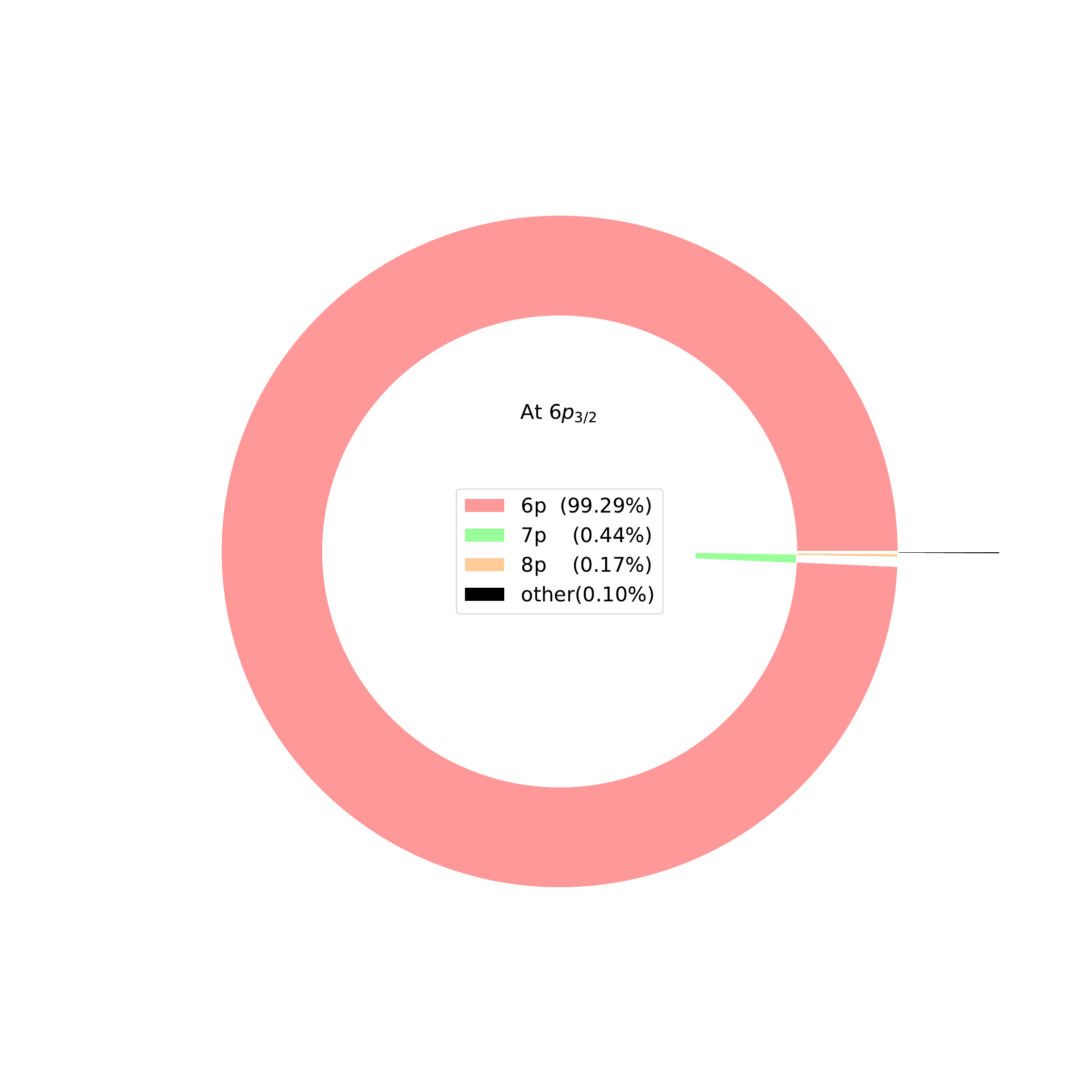}\\
	\end{tabular}
	\caption{Composition of X2C-ROHF valence $p_{1/2}$ and $p_{3/2}$ spinors of Br, I, and At in the basis of sf-X2C $p$ orbitals.}
	\label{overlap}
\end{figure}

As demonstrated in Table \ref{At_SOS} for the spin-orbit splittings (SOS) of the At $np$ orbitals,
the vtz-5p$^1_2$ basis set for At improves the vtz basis set dramatically, to the extent that
the SOSs for $2p$ to $8p$ described by the parent uncontracted basis set\cite{dyalltz} are completely reproduced.
It is for sure that the SOSs of even higher $p$ orbitals can also be described very well by inserting a few more projected scalar functions
to the vtz-5p$^1_2$ basis. The reason for us to stop at vtz-5p$^1_2$ lies in that
the At $6p_{1/2}$ spinor [calculated by X2C-ROHF (restricted open-shell Hartree-Fock)] is composed
mainly of $5p$ (0.41\%), $6p$ (98.13\%), $7p$ (0.38\%), and $8p$ (0.67\%), with marginal contributions
from $4p$ (0.04\%) and $9p$ (0.26\%). For comparison, the At $6p_{3/2}$ spinor is composed predominately of
$6p$ (99.29\%) and $7p$ (0.44\%). As such, the vtz-5p$^1_2$ basis set
is enough for the present purpose. On the other hand, the $np$ shells ($n \geq 10$; both scalar and spinor) by the vtz-5p$^1_2$ basis
do not have one-to-one correspondence with those by the uncontracted basis. Close inspections reveal that such high-lying virtual orbitals are composed mainly
of steep $p$ functions that also dominate the $2p$ and $3p$ shells of At. As such, they can be characterized as
 `anti-bonding' orbitals of the innermost orbitals, i.e., HELOs\cite{SOiCI}. Since the
innermost orbitals are to be kept frozen, such HELOs should also be removed from SOiCISCF calculations,
e.g., by means of the VSD approach\cite{SDSRev,SOiCI}. If the cc-pV5Z-PP (cc-pVTZ-PP) basis
\cite{ccpV5ZPP1,ccpV5ZPP2} is employed in the VSD, the $12p13p$ ($10p$-$13p$) shells will be removed from the orbital spectrum of the vtz-5p$^1_2$ basis.
It turns out that the two sets of calculations lead to very similar results. To be on the safe side, the cc-pV5Z-PP basis is always employed in the VSD.
Further in view of the compositions of Br $4p_{1/2}$ and I $5p_{1/2}$ (cf. Fig. \ref{overlap}), it appears that
a general rule of thumb for the description of $np_{1/2}$ is to
use $(n-n_o)p$ to $(n+n_v)p$ scalar orbitals, with $(n_o,n_v)=(0,1)$ and $(1,2)$ for $n\le 4$ and $n\ge 5$, respectively.

\begin{table}[!htp]
	\footnotesize
	\centering
	\caption{Spin-orbit splittings (SOS in a.u.) of the At $np$ orbitals by the sf-X2C+so-DKH1 and X2C Hamiltonians at the ROHF level }
	\begin{threeparttable}			
\begin{tabular}{c|cccccccc}\toprule
&\multicolumn{6}{c}{sf-X2C+so-DKH1} & \multicolumn{2}{c}{X2C}\\
&\cline{1-6}\\
&\multicolumn{2}{c}{vtz} &\multicolumn{2}{c}{vtz-$5$p$^1_2$\tnote{a}} & \multicolumn{2}{c}{UC\tnote{b}}
&\multicolumn{2}{c}{UC\tnote{b}}\\ 			
&   $p_{1/2}$ & SOS & $p_{1/2}$ & SOS & $p_{1/2}$ & SOS & $p_{1/2}$ & SOS \\\toprule
2p  &-616.0978 &88.3884 &-631.4414 	&101.5714 	 &-631.5769 	&101.7192     &-625.9119 	&98.5124 \\
3p	&-148.1337 &18.9563 &-152.9627 	&22.9141 	   &-152.9903 	&22.9486      &-151.9095 	&22.5089 \\
4p	&-35.3813  &4.5455  &-36.8270 	&5.6810 	   &-36.8324 		&5.6894       &-36.6611 	&5.7083  \\
5p	&-7.7718   &0.8579  &-8.0973 		&1.1056 	   &-8.0983 		&1.1072       &-8.1185 	  &1.1797  \\
6p	&-1.8419   &0.1320  &-1.8999 		&0.1758 	   &-1.9001 		&0.1760       &-1.8959 	  &0.1778  \\
\\
7p	&-0.8866   &0.0418  &-0.9021 		&0.0508 	   &-0.9022 		&0.0509       &-0.9013 	  &0.0518  \\
8p	&-0.4242   &0.0953  &-0.4492 		&0.0839 	   &-0.4495 		&0.0836       &-0.4483 	  &0.0805  \\
9p	&   -      &  -     &1.6210     &0.4886      &1.5355 		  &0.4438       &1.5435 	  &0.4555  \\
10p	&   -      &  -     &(10.9102)\tnote{c} 	&(2.1614)\tnote{c}    &8.8258 		  &1.4245       &8.8632 	  &1.4415  \\
11p	&   -      &  -     &(53.0826)\tnote{c} 	&(11.3610)\tnote{c}   &29.7183 		&3.9967       &29.8703 	  &3.9857  \\
12p &   -      &  -     &(297.6197)\tnote{c} &(79.4143)\tnote{c}   &84.6338     &10.2842      &85.1042    &10.1346 \\
13p &   -      &  -     &(2860.4246)\tnote{c}&(2119.9824)\tnote{c} &218.5297    &24.5449      &219.8068   &23.9551 \\
			\midrule
		\end{tabular}		
\begin{tablenotes}
\item[a] vtz augmented with 5 scalar functions projected out from the X2C $2p_{1/2}$ to $6p_{1/2}$ spinors.
\item[b] Uncontracted basis set\cite{dyalltz}.
\item[c] Spinors of no correspondence with those computed with the UC basis.
	\end{tablenotes}
\end{threeparttable}\label{At_SOS}
\end{table}

\subsection{SOSs of halogens} \label{ssec SOSs}

%


The results for the SOS of the At $^2P$ state are first discussed in detail.
The starting point is the minimal active space, CAS(5,3), which is equivalent to ROHF. As can be seen from Table \ref{Br2At},
the result is ca. -13\% away from  that\cite{sf-X2C-EOM-SOC2017} by
the Fock-space coupled-cluster with singles and doubles (FS-CCSD)\cite{visscher2001formulation} in conjunction with the molecular mean-field X2C
(mmf-X2C) Hamiltonian\cite{X2Cmmf2009}. The next try is to add in $7p$. As can be seen from the CAS(5,6) row in Table \ref{Br2At},
the SOiCISCF (= SOCASSCF) result is completely meaningless. It turns out that
orbital relaxations are in this case too much biased towards $^2P_{3/2}$, that is, $^2P_{3/2}$ is lowered much more than $^2P_{1/2}$,
thereby resulting in a terrible overestimate of the SOS.
Such situation can be avoided by further adding in the occupied $5p$ shell, i.e., CAS(11,9). This exercise
shows that the occupied and virtual orbitals must be treated in a balanced manner.
As a matter of fact, as already known from the composition of At $6p_{1/2}$,
the $5p$ to $8p$ shells must be included simultaneously, leading to CAS(11,12).
The result is indeed improved remarkably, with the error reduced to ca. -4\%.
Another support of this rule is that
further adding in $9p$ to $11p$ shells (i.e., CAS(11,15), CAS(11,18),
and CAS(11,21)) does not lead to discernible improvement (by less than 1\%).
On the other hand, the inclusion of the HELOs ($12p$ and $13p$ for At) destroys essentially all the calculations. For instance,
the CAS(11,12) result becomes 28295 cm$^{-1}$, which overestimates the mmf-X2C-FS-CCSD value by 21\% (see
Table S1 in the supporting information). All these point to that the present SOiCISCF-CAS(11,12) result should
agree with that by the spinor-based minimal active space X2C-CASSCF\cite{LixiaosongX2CCASSCF}.
However, this is not the case. Since the
difference between the sf-X2C+so-DKH1 and X2C Hamiltonians is really minor (cf. Table \ref{At_SOS}),
the large discrepancy between the present (22381cm$^{-1}$) and
their (19180 cm$^{-1}$) values should be ascribed to the use of different basis sets.
To check this, we carried out SOiCISCF-CAS(11,12) calculations with their basis set (segmented contracted Sapporo-DKH3-DZP-2012\cite{sapporo})
and the vtz-$5$p$^1_2$ set constructed from the uncontracted Sapporo-DKH3-DZP-2012 set according to the present prescription.
The results are 21325 and 22430 cm$^{-1}$, respectively. Both are closer to the present value than to theirs.
As such, the discrepancy between SOiCISCF and X2C-CASSCF cannot be resolved at the current stage.

Different from the direct involvement of $p$ shells in the expansions of $6p_{1/2}$ and $6p_{3/2}$,
the contributions from other shells are solely correlation effect, due to symmetry reasons. As can be seen from Table \ref{Br2At},
the addition of the $s$ and $d$ shells on top of CAS(11,12) does improve further the CAS(11,12) result.
However, the result becomes slightly worsened after further adding in the $5f$ shell (i.e., CAS(25,33)),
and the worsening is more enhanced when all the virtual orbitals other than the HELOs ($12p$ and $13p$)
are included simultaneously (i.e., CAS(25,47)). These findings reveal that
electron correlation and SOC reside in different regions of the Hilbert space
and may counteract each other. Nevertheless, SOiCISCF performs uniformlly better than SOiCI,
showing the importance of orbital relaxations for elements as heavy as At.

The above findings hold largely for Br and I as well. Therefore, the same active spaces
can be employed for other $p$-block elements.

\begin{table}[!htp]
	\scriptsize
	\centering
	\caption{The SOSs (in cm$^{-1})$ of Br, I, and At with the vtz-$m$p$^1_2$ basis sets. In parentheses
are the percentage deviations from experimental values }
	\begin{threeparttable}
		\centering
		\begin{tabular}{c|ccccccc}\toprule
			\multirow{2}{*}{$^2P_{1/2}$$-$$^{2}P_{3/2}$}&\multicolumn{2}{c}{Br}  & \multicolumn{2}{c}{I} & \multicolumn{2}{c}{At} &\multirow{2}{*}{Active orbital}\\
			& \multicolumn{1}{c}{SOiCI(2)} & SOiCISCF(2)
			& \multicolumn{1}{c}{SOiCI(2)} & SOiCISCF(2)
			& \multicolumn{1}{c}{SOiCI(2)} & SOiCISCF(2) &\\\toprule
CAS(5, 3)& 3424 (-7.09\%) &3424 (-7.09\%) &6993 (-8.02\%) &6993 (-8.02\%) &20272 (-13.35\%) &20272 (-13.35\%) &$np$\\
CAS(5, 6)&3456 (-6.21\%) &3481 (-5.53\%) &7090 (-6.75\%) &7545 (-0.76\%) &21487 (-8.15\%) &41041 (75.43\%)  & +$(n$+$1)p$\\
CAS(11, 9)   &3767 (2.22\%) &3746 (1.66\%) &7075 (-6.94\%) &7408 (-2.57\%) &21571 (-7.79\%) &20937 (-10.50\%) &+$(n$--$1)p$\\
CAS(11, 12) &3786 (2.75\%) &3773 (2.37\%) &7137 (-6.13\%) &7516 (-1.15\%) &21950 (-6.17\%) &22381 (-4.33\%) & +$(n$+$2)p$ \\
CAS(11, 15) &3760 (2.04\%) &3758 (1.98\%) &7387 (-2.83\%) &7339 (-3.47\%) &22055 (-5.72\%) &22432 (-4.11\%) & +$(n$+$3)p$\\
CAS(11, 18) &3787 (2.77\%) &3787 (2.77\%) &7369 (-3.07\%) &7382 (-2.90\%) &22077 (-5.63\%) &22509 (-3.78\%) & +$(n$+$4)p$\\
CAS(11, 21) &3739 (1.45\%) &3742 (1.55\%) &7398 (-2.70\%) &7403 (-2.63\%) &22532 (-3.69\%) &22541 (-3.65\%) &+$(n$+$5)p$\\
\\
CAS(13, 14)  &3817 (3.59\%) &3802 (3.18\%) &7208 (-5.19\%) &7570 (-0.44\%) &22059 (-5.71\%) &22479 (-3.91\%) & CAS(11, 12)+$ns(n$+1$)s$\\
CAS(15, 16) &3824 (3.78\%) &3809 (3.36\%) &7135 (-6.16\%) &7429 (-2.29\%) &21929 (-6.26\%) &22625 (-3.29\%) &+$(n$-1$)s(n$+2$)s$ \\
CAS(25, 26)&3678 (-0.19\%) &3775 (2.45\%) &7167 (-5.73\%) &7449 (-2.02\%) &20833 (-10.95\%) &23000 (-1.69\%) &  +$(n$-1$)dnd$\\
CAS(25, 33)  &3667 (-0.50\%) &3764 (2.15\%) &7287 (-4.15\%) &7511 (-1.22\%) &20684 (-11.58\%) &22735 (-2.82\%) & +$(n$-1$)f$\\
\\
CAS(25, 31)&3594 (-2.46\%) &3692 (0.20\%) &7075 (-6.94\%) &7122 (-6.33\%) &20472 (-12.49\%) &22633 (-3.25\%) &  CAS(25, 26)+$(n$+1$)d$\\
CAS(25, 40)  &3622 (-1.71\%) &3621 (-1.73\%) &7040 (-7.41\%) &7038 (-7.42\%) &22057 (-5.72\%) &22720 (-2.88\%) & +$(n$+3$)p(n$+4$)p(n$+5$)p$\\
CAS(25, full) &3636 (-1.32\%) &3637 (-1.30\%) &7209 (-5.18\%) &7211 (-5.16\%) &21724 (-7.14\%) &22172 (-5.22\%) & +$(n$-1$)f$\\
Extrapolated\tnote{a} &3636 (-1.37\%) &3636 (-1.34\%) &7003 (-7.90\%) &7002 (-7.90\%) &22150 (-5.32\%) &22257 (-4.86\%) & -\\
			Exp. &\multicolumn{2}{c}{3685\tnote{b}}	&	\multicolumn{2}{c}{7603\tnote{a}}	&	\multicolumn{2}{c}{23394\tnote{c}} &  \\\midrule
		\end{tabular}
\begin{tablenotes}
\item[a] Extrapolated CAS(25,full) result (linear fit of the results by using $C_{\mathrm{min}}=\{25, 10, 7.5\}\times 10^{-6}$).
\item[b] NIST\cite{NIST1}.
\item[c] mmf-X2C-FS-CCSD result\cite{sf-X2C-EOM-SOC2017}.
	\end{tablenotes}
	\end{threeparttable}\label{Br2At}

\end{table}

\subsection{SOC-induced orbital relaxation}
\begin{figure}[!htp]
	\centering
	\begin{tabular}{c}
		\includegraphics[width=1.0\textwidth]{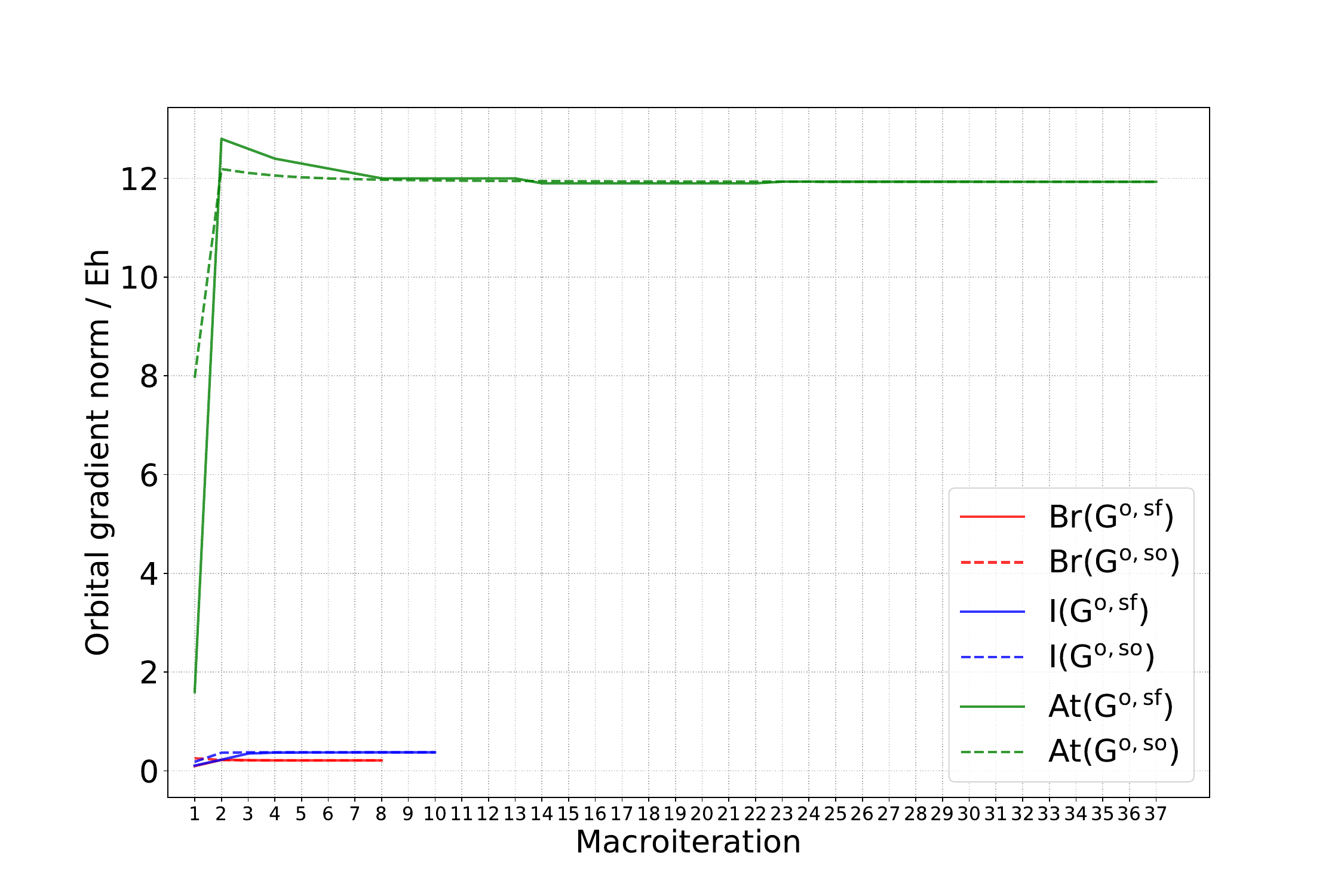}\\
		\\
	\end{tabular}
	\caption{Norms of spin-free (sf) and spin-dependent (so) orbital gradients of SOiCISCFCAS(13,14) calculations of  Br, I, and At.}
	\label{BrIAtgrad}
\end{figure}

\begin{figure}[!htp]
	\centering
	\begin{tabular}{cc}
		\includegraphics[width=1.0\textwidth]{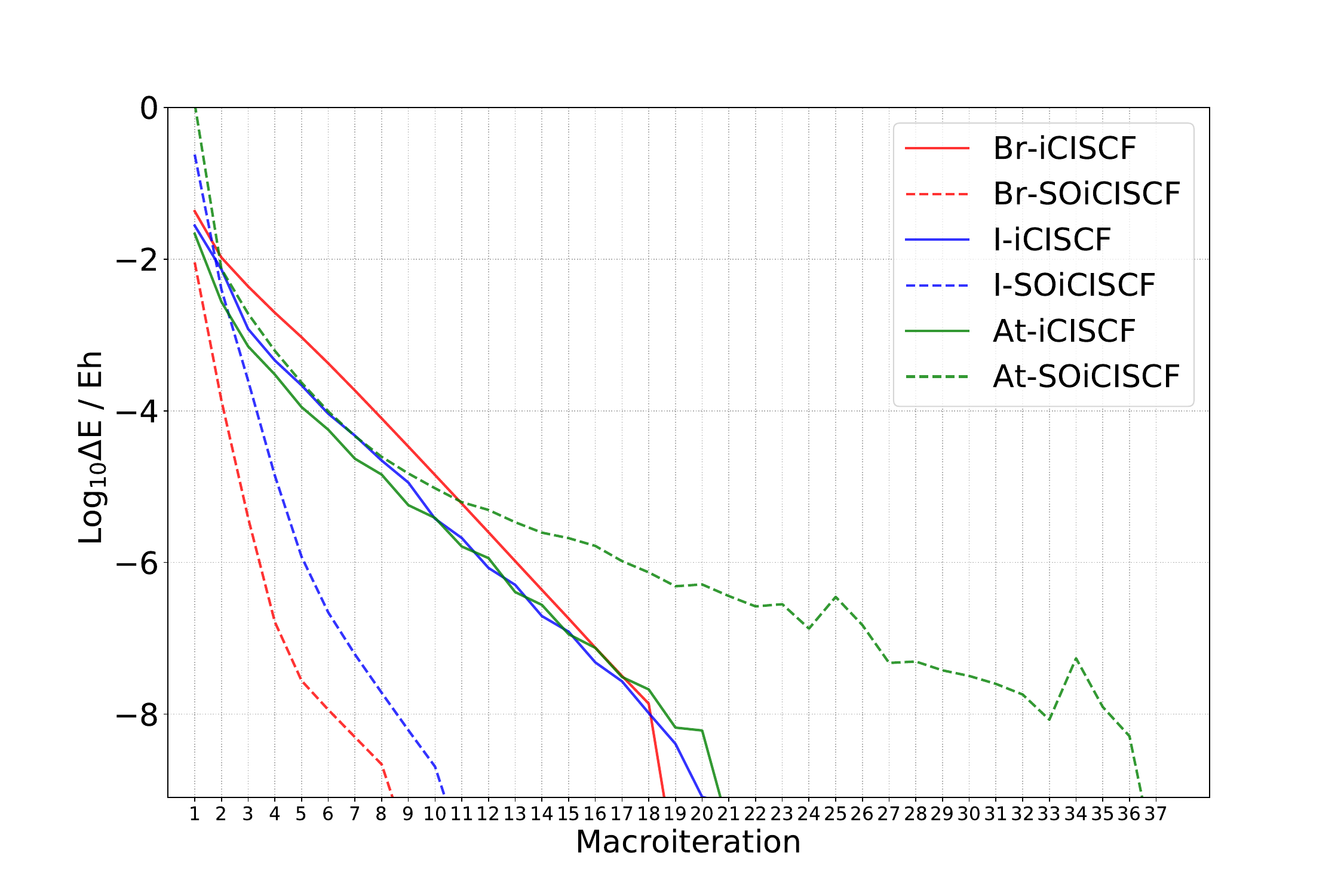}
	\end{tabular}
	\caption{Convergence patterns of iCISCF-CAS(13,14) and SOiCISCF-CAS(13,14) calculations of Br, I, and At.}
	\label{conv}
\end{figure}

To reveal the SOC-induced orbital relaxation effects in more detail, the norms of the orbital gradients in the
SOiCISCF-CAS(13,14) calculations of Br, I, and At are further plotted in Fig. \ref{BrIAtgrad} along the iterations.
It can be seen that the spin-dependent orbital gradient $G^{\mathrm{o,so}}$ of At is much larger than those of Br or I.
Since the spin-free orbital gradient $G^{\mathrm{o,sf}}$ is just the negative of $G^{\mathrm{o,so}}$ upon convergence
(cf. Eq. \eqref{fullEOa}), the sf-X2C-iCISCF orbitals of At must be deformed significantly to
counterbalance the large $G^{\mathrm{o,so}}$. This explains why the SOiCISCF-CAS(13,14) calculation of At
converges much slower than those of Br and I (cf. Fig. \ref{conv}).


\begin{figure}[!htp]
	\centering
	\begin{tabular}{c}
		\includegraphics[width=1.0\textwidth]{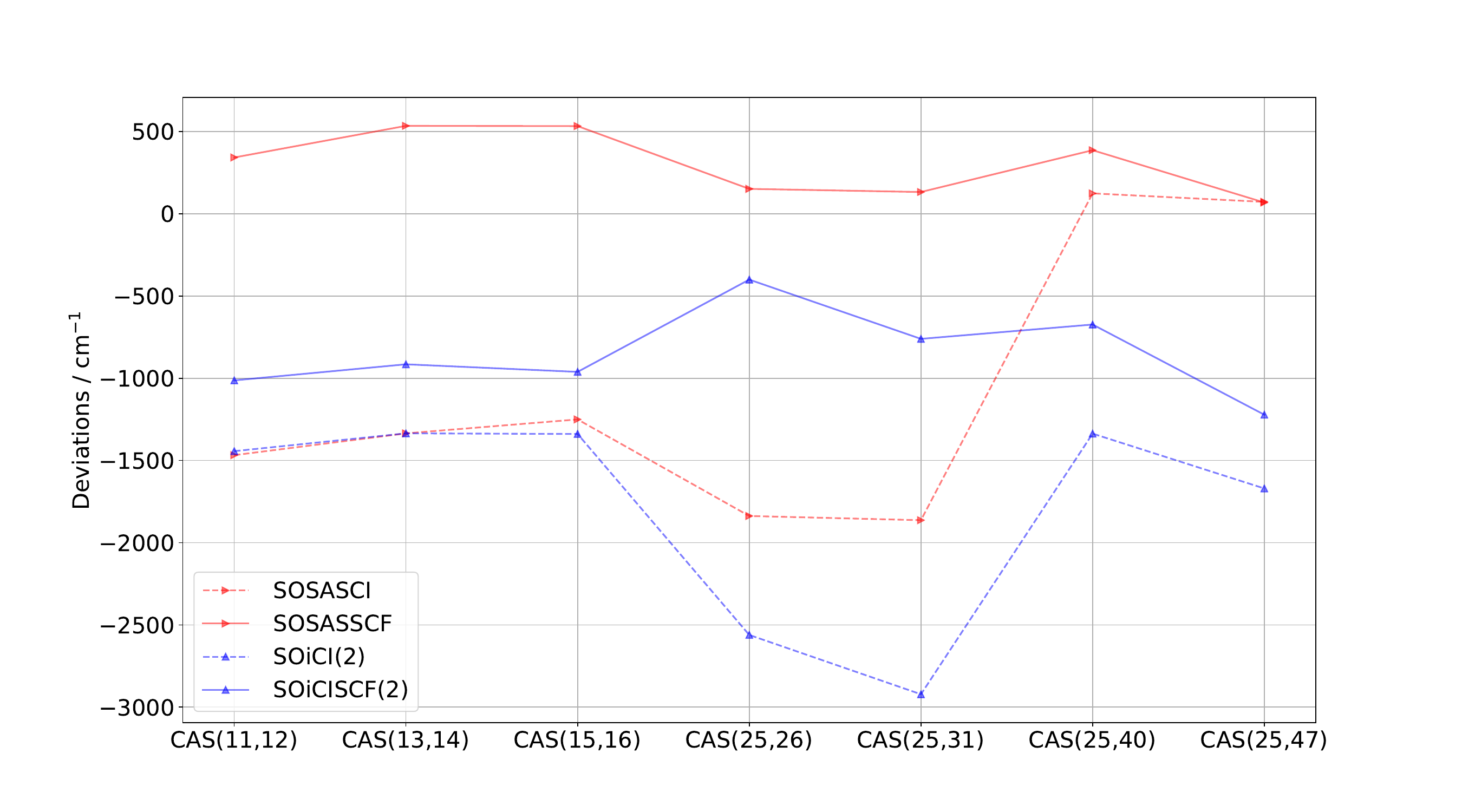}\\
	\end{tabular}
	\caption{Deviations of the SOSs of At by SOSASCI, SOSASSCF, SOiCI(2), and SOiCISCF(2) from mmf-X2C-FSCCSD. }
	\label{SOSASSCF}
\end{figure}

To separate electron correlation from orbital relaxation, we further carried out calculations
where the active spaces are composed only of the ROHF reference and single excitations from it
to the active orbitals. Due to symmetry reasons, such single excitations are purely $p\rightarrow p$ or $p\rightarrow f$.
The results are denoted as SOSASSCF when the orbitals are optimized and as SOSASCI when the orbitals are not optimized.
The difference between SOSASSCF and SOSASCI is then purely orbital relaxation, whereas that
between SOiCISCF(2) and SOiCI(2) contains both orbital relaxation and electron correlation.
As can be seen from Fig. \ref{SOSASSCF}, SOSASSCF is more accurate than SOSASCI for all active spaces, due to the inclusion of orbital relaxation effects.
Not surprisingly, when all virtual orbitals are included in the active space (i.e., CAS(25,47)), SOSASCI and SOSASSCF become virtually identical.
For the same reason, SOiCISCF(2) performs uniformly better than SOiCI(2). However, that SOSASSCF performs better than SOiCISCF(2)
cannot be taken seriously, since electron correlation
should always be taken into account. It is just that electron correlation and SOC should be treated in a balanced manner in SOiCISCF(2).

\subsection{SOSs of selected $\textit{p}$-block elements}
\begin{figure}[!htp]
	\centering
	\begin{tabular}{c}
		\includegraphics[width=1.0\textwidth]{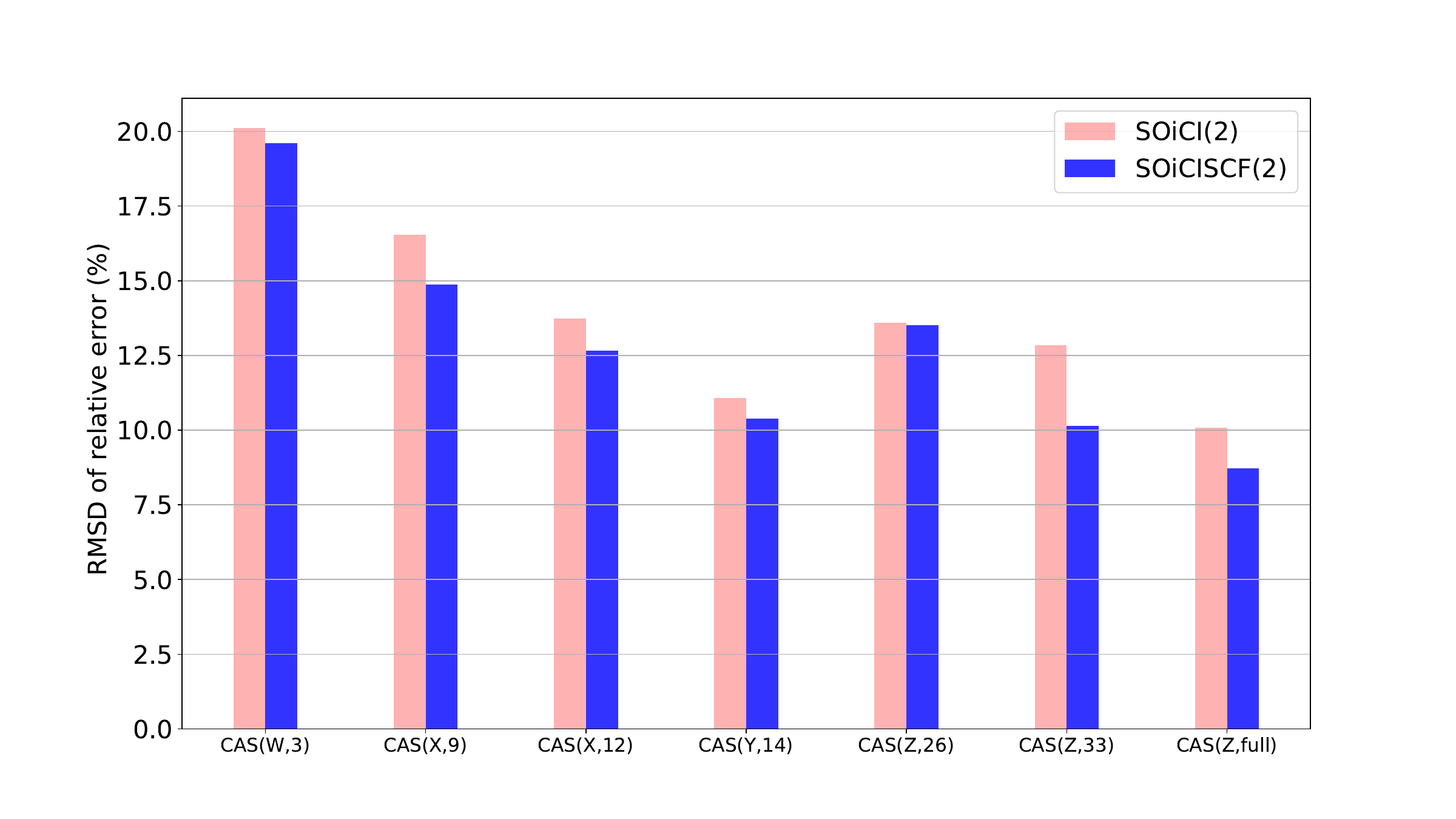}\\
	\end{tabular}
	\caption{Percentage root mean-square deviations (RMSD) of all SOSs of the fourth- to sixth-row $p$-block elements computed by SOiCI(2) and SOiCISCF(2) with different active spaces. The numbers of active electrons are denoted as W=[1, 5], X=[7, 11], Y=[9, 13], and Z=[21, 25], which are dependent on the groups.}
	\label{RMSD}
\end{figure}

The SOSs of the ground and low-lying excited states of the fourth- to sixth-row $p$-block elements were all calculated
using SOiCI(2) and SOiCISCF(2).
Since there exist no new mechanisms concerning SOC-induced orbital relaxations (the primary focus here), we decide to
put the results in the Supporting Information (see Tables S2-S5) and make only a gross comparison with the experimental values\cite{NIST1}.
The percentage root mean-square deviations (RMSD) of
all calculated SOSs are shown in Fig. \ref{RMSD} for different active spaces (see Sec. \ref{ssec SOSs}).
It can be concluded that SOiCISCF(2) outperforms SOiCI(2) for the same active space. However,
the results also show that the convergence with respect to active space size is not monotonic, resulting again from
the counterplay between (static) electron correlation and SOC. The situation should be improved when dynamic correlation
is further accounted for.

\begin{figure}[!htp]
	\centering
	\begin{tabular}{c}
		\includegraphics[width=1.0\textwidth]{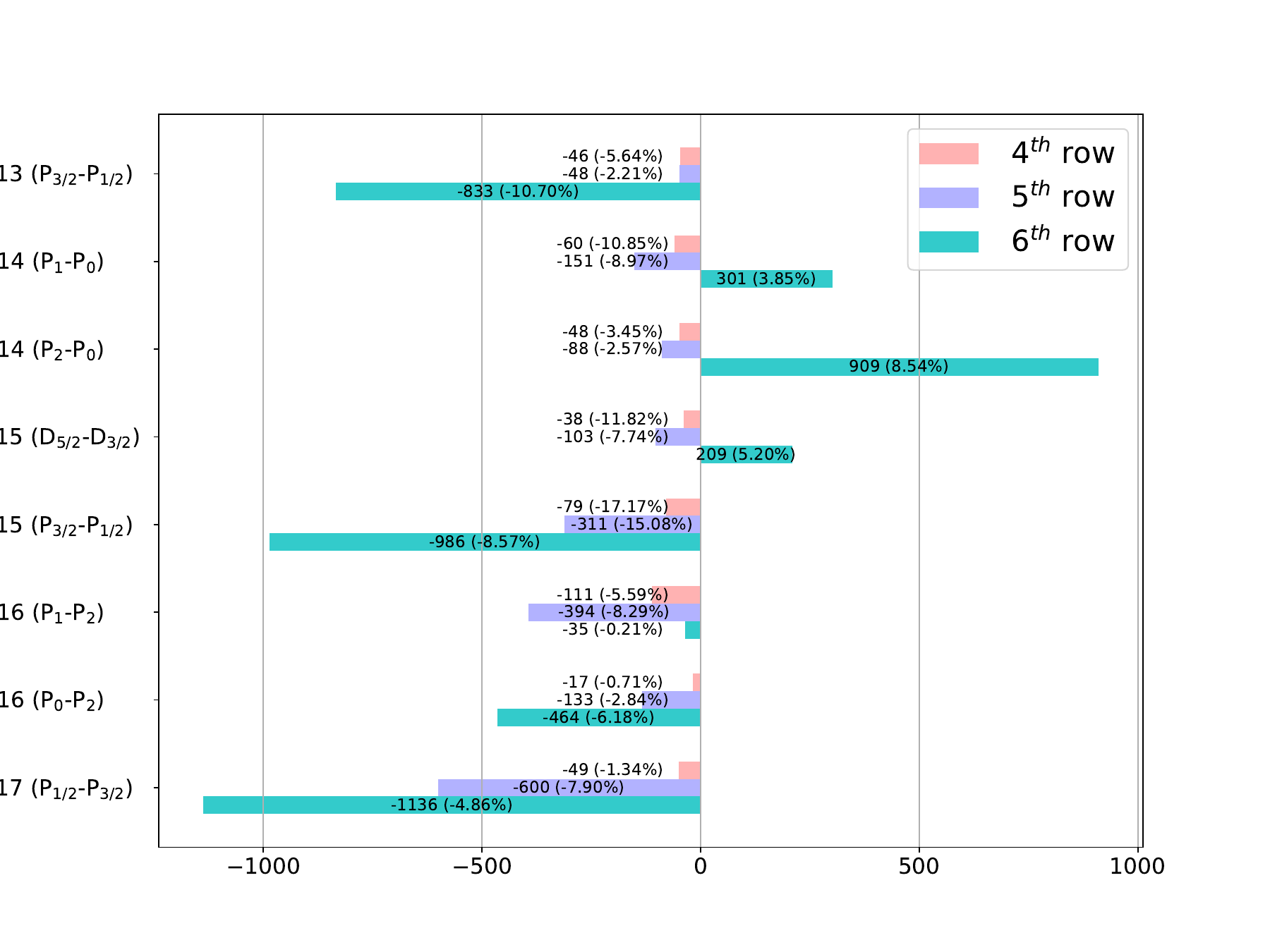}\\
	\end{tabular}
	\caption{Errors (in cm$^{-1}$) of the extrapolated SOiCISCF SOSs for groups 13 to 17 $p$-block elements.
In parentheses are percentage errors from the experimental values.}
	\label{extrapolated}
\end{figure}

Finally, when all virtual MOs are included in the active spaces, the energies of the spinor states were further extrapolated linearly
using three $C_{\mathrm{min}}$ values, \{25, 10, 7.5\}$\times$10$^{-6}$. The errors of the so-extrapolated SOSs
from the experimental values\cite{NIST1} are plotted in Fig. \ref{extrapolated}. It can be seen that SOiCISCF(2)
performs very well for the fourth- and fifth-row $p$-block elements (with the largest error being ca. 600 cm$^{-1}$)
but deteriorates for the sixth-row $p$-block elements. Still, however, the largest error therein is less than 11\%.
It remains to see whether the situation will be improved when the third-order so-DKH Hamiltonian\cite{X2CSOCBook2017,X2CSOC1,X2CSOC2}
is employed. Yet, the corresponding third-order two-body so-DKH operator should first be worked out.
Anyhow, the present SOiCISCF(2) approach can safely be applied to molecular systems containing even $6p$-block elements,
where SOCs are quenched to some extent by chemical bondings\cite{sf-X2C-EOM-SOC2017}.

\section{Conclusions and Outlook}\label{Conclusion}
A relativistic multiconfiguration self-consistent field approach has been proposed to treat both scalar relativity and SOC variationally, yet with
the orbitals kept real-valued. It is a natural combination of the sf-iCISCF (solver of large CAS)
and SOiCI (equal treatment of correlation and SOC) approaches developed by our group and is thus dubbed as SOiCISCF.
Compared to SOiCI, SOiCISCF can capture all SOC-induced orbital relaxations, as demonstrated by pilot calculations of $p$-block elements.
Apart from this, the variational treatment of SOC allows an easy treatment of physical properties and analytic energy gradients in the presence of SOC.
Yet, a drawback of SOiCISCF (and other similar methods) should be noted: The counterplay between correlation and SOC often
renders the result by a larger active space worse than that by a smaller one. However, this can be remedied by further accounting
for dynamic correlation, where real-valued orbitals are much favored over complex-valued spinors. Work along these directions
are undertaken at our laboratory.

\section*{Acknowledgment}
This work was supported by the National Natural Science Foundation of China (Grant Nos. 21833001, 21973054, and 22273052),
Mountain Tai Climb Program of Shandong Province, and Qilu Young Scholar Program of Shandong University.
YG is grateful to Profs. Y. Lei, Y. Zhang and F. Wang for helpful discussions.

\appendix

\section{Orbital Gradient and Hessian}\label{AppGRDHESS}
The following conventions are to be used throughout:
(1) the CI coefficients are generally complex but the orbitals are enforced to be real-valued;
(2) core, active, virtual, and arbitrary MOs are designated by $\{i,j,k,l\}$, $\{e,f,t,u,v,w\}$, $\{a,b,c,d\}$, and $\{p,q,r,s,x,y,z\}$, respectively;
(3) repeated indices are always summed up.
\subsection{Spin-free Case}
\setcounter{equation}{0}
\renewcommand{\theequation}{A\arabic{equation}}

The following commutators
\begin{align}
[a_{q\sigma},\hat{E}_{rs}]&=\delta_{qr}a_{s\sigma},\\
[a_{q\sigma},\hat{e}_{rs,xy}]&=\delta_{qr}\hat{E}_{xy}a_{s\sigma}+\delta_{qx}\hat{E}_{rs}a_{y\sigma},\\
[a^{p\sigma},\hat{E}_{rs}]&=-\delta_{ps}a^{r\sigma},\\
[a^{p\sigma},\hat{E}_{rs,xy}]&=-[a_{p\sigma},\hat{E}_{sr,yx}]^\dag=-\delta_{ps}a^{r\sigma}\hat{E}_{xy}-\delta_{py}a^{x\sigma}\hat{E}_{rs},\\
[\hat{E}_{pq}, \hat{E}_{rs}]&=\delta_{qr}\hat{E}_{ps}-\delta_{ps}\hat{E}_{rq}, \label{Commutator}\\
[\hat{E}_{pq}, \hat{e}_{rs,xy}]&=\delta_{qr}\hat{e}_{ps,xy}-\delta_{ps}\hat{e}_{rq,xy}+\delta_{qx}\hat{e}_{py,rs}-\delta_{py}\hat{e}_{xq,rs},\\
0&=[\hat{A},[\hat{B},\hat{C}]]+[\hat{B},[\hat{C},\hat{A}]]+[\hat{C},[\hat{A},\hat{B}]]\label{JacobiID}
\end{align}
 can be used repeatedly to obtain the following expressions
\begin{align}
[\hat{E}_{pq}^-,\hat{E}_{rs}]&=(1-\delta_{pq})(1-P_{pq})[\hat{E}_{pq},\hat{E}_{rs}]=(1-\delta_{pq})(1-P_{pq})(\delta_{qr}\hat{E}_{ps}-\delta{ps}\hat{E}_{rq}),\\
[a_{q\sigma},\hat{H}^{\mathrm{sf}}]&=h^{\mathrm{sf}}_{qz}a_{z\sigma}+g_{qz,xy}\hat{E}_{xy}a_{z\sigma},\label{Fhalf}\\
[a^{p\sigma},\hat{H}^{\mathrm{sf}}]&=-[a_{p\sigma},\hat{H}^{\mathrm{sf}}]^\dag=-h^{\mathrm{sf}}_{zp}a^{z\sigma}-g_{zp,xy}a^{z\sigma}\hat{E}_{xy},\\
\hat{F}_{pq}^{\mathrm{sf}}&=a^{p\sigma}[a_{q\sigma}, \hat{H}^{\mathrm{sf}}]=\hat{E}_{pz}h^{\mathrm{sf}}_{qz}+\hat{e}_{pz,xy}g_{qz,xy}, \label{Koopmans-op}\\
\hat{\tilde{F}}^{\mathrm{sf}}_{pq}&=[a^{p\sigma}, \hat{H}^{\mathrm{sf}}]a_{q\sigma}=-(a^{q\sigma}[a_{p\sigma}, \hat{H}^{\mathrm{sf}}])^\dag
=-(\hat{F}^{\mathrm{sf}}_{qp})^\dag=-P_{pq}h.c.\hat{F}^{\mathrm{sf}}_{pq},\\
F_{pq}^{\mathrm{sf}}&=\langle 0|\hat{F}_{pq}^{\mathrm{sf}}|0\rangle=D_{pz}h^{\mathrm{sf}}_{qz}+\Gamma_{pz,xy}g_{qz,xy},\label{sfGenFock}\\
\tilde{F}_{pq}^{\mathrm{sf}}&=\langle 0|\hat{\tilde{F}}^{\mathrm{sf}}_{pq}|0\rangle
=-P_{pq}\langle 0|(\hat{F}^{\mathrm{sf}}_{pq})^\dag|0\rangle=-P_{pq}\langle 0|\hat{F}^{\mathrm{sf}}_{pq}|0\rangle^*=-P_{pq}F_{pq}^{\mathrm{sf}*}\nonumber\\
&=-\hat{E}_{zq}h^{\mathrm{sf}}_{zp}-\hat{e}_{zq,xy}g_{zp,xy},\\
[\hat{E}_{pq},\hat{H}^{\mathrm{sf}}]&=\hat{F}^{\mathrm{sf}}_{pq}-P_{pq}(\hat{F}^{\mathrm{sf}}_{pq})^\dag\\
&=\{\hat{E}_{pz}h^{\mathrm{sf}}_{qz}+\hat{e}_{pz,xy}g_{qz,xy}\}-\{\hat{E}_{zq}h^{\mathrm{sf}}_{zp}+\hat{e}_{zq,xy}g_{zp,xy}\},\\
[\hat{E}_{pq}^-,\hat{H}^{\mathrm{sf}}]&=[\hat{E}_{pq},\hat{H}^{\mathrm{sf}}]+h.c.=(1-\delta_{pq})(1-P_{pq})(\hat{F}^{\mathrm{sf}}_{pq}+h.c.)\label{GRDop}\\
&=(1-\delta_{pq})(1-P_{pq})\{(\hat{E}_{pz}h^{\mathrm{sf}}_{qz}+\hat{e}_{pz,xy}g_{qz,xy})+h.c.\},\label{Epq-Hsf}\\
%
%
[\hat{E}_{pq}^-,\hat{F}_{rs}^{\mathrm{sf}}]&=(1-\delta_{pq})(1-P_{pq})[\hat{E}_{pq},\hat{F}^{\mathrm{sf}}_{rs}]\\
&=(1-\delta_{pq})(1-P_{pq})\{\delta_{qr}\hat{F}^{\mathrm{sf}}_{ps}-(\hat{E}_{rq}h^{\mathrm{sf}}_{sp}+\hat{e}_{rq,xy}g_{sp,xy})\nonumber\\
&+\hat{e}_{rx,py}g_{sx,qy}-\hat{e}_{rx,yq}g_{sx,yp}\}\\
&=(1-\delta_{pq})(1-P_{pq})\{(\hat{E}_{rp}h^{\mathrm{sf}}_{sq}+\hat{e}_{rp,xy}g_{sq,xy})-\delta_{pr}\hat{F}^{\mathrm{sf}}_{qs}\nonumber\\
&+(\hat{e}_{px,ry}+\hat{e}_{xp,ry})g_{qx,ys}\}.\label{EFrs}
\end{align}

Note in passing that Eqs. \eqref{Fhalf} to \eqref{EFrs} hold even when the orbitals are complex.
For real orbitals considered here, the spin-free orbital gradient \eqref{Orbgrd} can be obtained from
Eq. \eqref{GRDop} as
\begin{align}
G^{\mathrm{o,sf}}_{pq}&=\langle 0|[\hat{E}_{pq}^-,\hat{H}^{\mathrm{sf}}]|0\rangle=G^{\mathrm{o,sf}*}_{pq}=-G^{\mathrm{o,sf}}_{qp}\\
&=2(1-\delta_{pq})(1-P_{pq})\Re F_{pq}^{\mathrm{sf}}.\label{sfOrbGrd}
\end{align}
To calculate the spin-free orbital Hessian, we first invoke Eqs. \eqref{GRDop} and \eqref{EFrs} to obtain
\begin{align}
[\hat{E}_{pq}^-,[\hat{E}_{rs}^-,\hat{H}^{\mathrm{sf}}]]&=(1-\delta_{pq})(1-\delta_{rs})(1-P_{rs})([\hat{E}_{pq}^-,\hat{F}_{rs}^{\mathrm{sf}}]+h.c.)\\
&=2(1-\delta_{pq})(1-\delta_{rs})(1-P_{pq})(1-P_{rs})\nonumber\\
&\times \{\frac{1}{2}(\hat{E}_{pr}+\hat{E}_{rp})h^{\mathrm{sf}}_{qs}-\frac{1}{2}\delta_{pr}(\hat{F}^{\mathrm{sf}}_{qs}+h.c.)
+\hat{P}^{xy}_{pr}J^{xy}_{qs}+2\hat{Q}^{xy}_{pr}K^{xy}_{qs}\},\label{Hessianopnosym}
\end{align}
where
\begin{align}
\hat{P}^{xy}_{pq}&=\frac{1}{4}(\hat{e}_{pq,xy}+\hat{e}_{qp,xy}+\hat{e}_{pq,yx}+\hat{e}_{qp,yx}),\\
\hat{Q}^{xy}_{pq}&=\frac{1}{4}(\hat{e}_{px,yq}+\hat{e}_{xp,yq}+\hat{e}_{px_,qy}+\hat{e}_{xp,qy}),\\
J^{xy}_{pq}&=g_{pq,xy}=J^{yx}_{pq}=J^{xy}_{qp}=J^{pq}_{xy}=K^{px}_{qy},\\
K^{xy}_{pq}&=g_{px,yq}=K^{yx}_{qp}=K^{pq}_{xy}=K^{qp}_{yx}=J^{px}_{qy}.
\end{align}
We then have the following fully symmetrized orbital Hessian operator
\begin{align}
\hat{E}^{\mathrm{oo,sf}}_{pq,rs}&=\frac{1}{2}(1+P_{pr}P_{qs})[\hat{E}_{pq}^-,[\hat{E}_{rs}^-,\hat{H}^{\mathrm{sf}}]]\\
&=2(1-\delta_{pq})(1-\delta_{rs})(1-P_{pq})(1-P_{rs})\nonumber\\
&\times \{\frac{1}{2}(\hat{E}_{pr}+\hat{E}_{rp})h^{\mathrm{sf}}_{qs}-\frac{1}{4}\delta_{pr}[(\hat{F}^{\mathrm{sf}}_{qs}+\hat{F}^{\mathrm{sf}}_{sq})+h.c.]
+\hat{P}^{xy}_{pr}J^{xy}_{qs}+2\hat{Q}^{xy}_{pr}K^{xy}_{qs}\}.\label{Hessianop}
\end{align}
Since the above operator is manifestly Hermitian, its expectation (orbital Hessian) is real-valued, viz.,
\begin{align}
E^{\mathrm{oo,sf}}_{pq,rs}&=\langle 0|\hat{E}^{\mathrm{oo,sf}}_{pq,rs}|0\rangle=-E^{\mathrm{oo,sf}}_{qp,rs}=-E^{\mathrm{oo,sf}}_{pq,sr}=E^{\mathrm{oo,sf}}_{qp,sr}=E^{\mathrm{oo,sf}*}_{pq,rs}\nonumber\\
&=2(1-\delta_{pq})(1-\delta_{rs})(1-P_{pq})(1-P_{rs})\nonumber\\
&\times \{\gamma_{pr}h^{\mathrm{sf}}_{qs}-\frac{1}{2}\delta_{pr}\Re (F_{qs}^{\mathrm{sf}}+F_{sq}^{\mathrm{sf}})+P^{xy}_{pr}J^{xy}_{qs}+2Q^{xy}_{pr}K^{xy}_{qs}\}.\label{sfEOO}
\end{align}
where
\begin{align}
\gamma_{pq}&=\frac{1}{2}(D_{pq}+D_{qp})=\Re D_{pq},\\
P^{xy}_{pq}&=\frac{1}{4}(\Gamma_{pq,xy}+\Gamma_{qp,xy}+\Gamma_{pq,yx}+\Gamma_{qp,yx})\\
&=\frac{1}{2}\Re ((\Gamma_{pq,xy}+\Gamma_{pq,yx})=P_{pq}^{yx}=P^{xy}_{qp}=P^{pq}_{xy}=Q^{px}_{qy},\\
Q^{xy}_{pq}&=\frac{1}{4}(\Gamma_{px,yq}+\Gamma_{xp,yq}+\Gamma_{px,qy}+\Gamma_{xp,qy})\\
&=\frac{1}{2}\Re(\Gamma_{px,yq}+\Gamma_{px,qy})=Q^{yx}_{qp}=Q^{pq}_{xy}=Q^{qp}_{yx}=P^{px}_{qy}.
\end{align}
The diagonal term of $E^{\mathrm{oo,sf}}_{pq,rs}$ \eqref{sfEOO} can be calculated as
\begin{align}
E^{\mathrm{oo,sf}}_{pq,pq}&=\delta_{pr}\delta_{qs}E^{\mathrm{oo,sf}}_{pq,rs}\\
&=2(1-\delta_{pq})\{\gamma_{qq}h_{pp}^{\mathrm{sf}}+\gamma_{pp} h_{qq}^{\mathrm{sf}}-F_{pp}^{\mathrm{sf}}-F_{qq}^{\mathrm{sf}}\nonumber\\
&+P^{xy}_{pp}J^{xy}_{qq}+2Q^{xy}_{pp}K^{xy}_{qq}+P^{xy}_{qq}J^{xy}_{pp}+2Q^{xy}_{qq}K^{xy}_{pp}\nonumber\\
&-2(\gamma_{pq}h_{qp}^{\mathrm{sf}}+P^{xy}_{pq}J^{xy}_{qp}+2Q^{xy}_{pq}K^{xy}_{qp})\}\label{DiagOO}\\
&=2(1-\delta_{pq})\{(1+P_{pq})(-F^{\mathrm{sf}}_{qq}+\gamma_{pp}h^{\mathrm{sf}}_{qq}+P^{xy}_{pp}J^{xy}_{qq}+2Q^{xy}_{pp}K^{xy}_{qq})\nonumber\\
&-2(\gamma_{pq}h^{\mathrm{sf}}_{qp}+P^{xy}_{pq}J^{xy}_{qp}+2Q^{xy}_{pq}K^{xy}_{qp})\}.\label{sfDiagEOO}
\end{align}
Note in passing that Eq. \eqref{DiagOO} can only be obtained by acting $\delta_{pr}\delta_{qs}$ on the fully permuted expression of Eq. \eqref{sfEOO}.

Eq. \eqref{sfEOO} can be obtained in an alternative manner\cite{ElectronStructure}, by starting with the following identity
\begin{align}
\langle 0|[\hat{E}_{pq},[\hat{E}_{rs},\hat{H}^{\mathrm{sf}}]]|0\rangle&=\langle 0|a^{p\sigma}[a_{q\sigma},[\hat{E}_{rs},\hat{H}^{\mathrm{sf}}]]|0\rangle
+\langle 0|a^{q\sigma}[a_{q\sigma},[\hat{E}_{sr},\hat{H}^{\mathrm{sf}}]]|0\rangle^*\\
&=G_{pq,rs}+G_{qp,sr}^*=(1+P_{pq}P_{rs}c.c.)G_{pq,rs},
\end{align}
where $G_{pq,rs}$ can be split into three terms in view of the Jacobi identity \eqref{JacobiID}:
\begin{align}
G_{pq,rs}&=\langle 0|a^{p\sigma}[a_{q\sigma},[\hat{E}_{rs},\hat{H}^{\mathrm{sf}}]]|0\rangle=A_{pq,rs}+B_{pq,rs}+C_{pq,rs},\label{Jacobi2}\\
A_{pq,rs}&=\langle 0|a^{p\sigma}[[a_{q\sigma},\hat{E}_{rs}],\hat{H}^{\mathrm{sf}}]|0\rangle
=\delta_{qr}\langle 0|a^{p\sigma}[a_{s\sigma},\hat{H}^{\mathrm{sf}}]|0\rangle
=\delta_{qr}F_{ps}^{\mathrm{sf}},\\
B_{pq,rs}&=\langle 0|a^{p\sigma}[\hat{E}_{rs},[a_{q\sigma},\hat{h}^{\mathrm{sf}}]]|0\rangle,\label{Bdef}\\
C_{pq,rs}&=\langle 0|a^{p\sigma}[\hat{E}_{rs},[a_{q\sigma},\hat{g}^{\mathrm{sf}}]]|0\rangle.\label{Cdef}
\end{align}
Eqs. \eqref{Bdef} and \eqref{Cdef} can be calculated via Eq. \eqref{Fhalf} and following identities
\begin{align}
a^{p\sigma}[\hat{E}_{rs},a_{z\sigma}]&=-\delta_{rz}E_{ps},\\
a^{p\sigma}[\hat{E}_{rs},\hat{E}_{xy}a_{z\sigma}]&=\delta_{sx}a^{p\sigma r\tau}_{z\sigma y\tau}-\delta_{ry}a^{p\sigma x\tau}_{z\sigma s\tau}
-\delta_{rz}a^{p\sigma x\tau}_{s\sigma y\tau},
\end{align}
such that
\begin{align}
B_{pq,rs}&=-D_{ps}h^{\mathrm{sf}}_{qr},\\
C_{pq,rs}&=\Gamma_{pz,ry}g_{qz,sy}-\Gamma_{pz,xs}g_{qz,xr}-\Gamma_{ps,xy}g_{qr,xy}\\
&=\Gamma_{px,ry}g_{qx,sy}-\Gamma_{px,ys}g_{qx,yr}-\Gamma_{ps,xy}g_{qr,xy}.
\end{align}
Further in view of the identity $P_{pq}^2=1$ and hence
\begin{align}
(1-P_{pq})(1-P_{rs})P_{pq}P_{rs}=(1-P_{pq})(1-P_{rs}),
\end{align}
we obtain
\begin{align}
&(1-P_{pq})(1-P_{rs})(G_{pq,rs}+P_{pq}P_{rs}G_{pq,rs}^*)=(1-P_{pq})(1-P_{rs})(G_{pq,rs}+G_{pq,rs}^*)\\
&=(1-P_{pq})(1-P_{rs})\{-(D_{ps}+D_{sp})h^{\mathrm{sf}}_{qr}+\delta_{qr}(F_{ps}^{\mathrm{sf}}+F_{ps}^{\mathrm{sf}*})\nonumber\\
&+\Gamma_{px,ry}g_{qx,sy}-\Gamma_{px,ys}g_{qx,yr}-\Gamma_{ps,xy}g_{qr,xy}\nonumber\\
&+\Gamma_{xp,yr}g_{qx,sy}-\Gamma_{xp,sy}g_{qx,yr}-\Gamma_{sp,xy}g_{qr,xy}\}\\
&=(1-P_{pq})(1-P_{rs})\{(D_{pr}+D_{rp})h^{\mathrm{sf}}_{qs}-\delta_{pr}(F^{\mathrm{sf}}_{qs}+F_{qs}^{\mathrm{sf}*})\nonumber\\
&+(\Gamma_{px,yr}+\Gamma_{px,ry}+\Gamma_{xp,ry}+\Gamma_{xp,yr})g_{qx,ys}\nonumber\\
&+(\Gamma_{pr,xy}+\Gamma_{rp,xy})g_{qs,xy}\}.
\end{align}
Therefore,
\begin{align}
E^{\mathrm{oo,sf}}_{pq,rs}&=\frac{1}{2}(1-\delta_{pq})(1-\delta_{rs})(1+P_{pr}P_{qs})(1-P_{pq})(1-P_{rs})(G_{pq,rs}+G_{pq,rs}^*)\\
&=2(1-\delta_{pq})(1-\delta_{rs})(1-P_{pq})(1-P_{rs})\nonumber\\
&\times \{\gamma_{pr}h^{\mathrm{sf}}_{qs}-\frac{1}{2}\delta_{pr}\Re (F_{qs}^{\mathrm{sf}}+F_{sq}^{\mathrm{sf}})+P^{xy}_{pr}J^{xy}_{qs}+2Q^{xy}_{pr}K^{xy}_{qs}\},\label{sfEOOb}
\end{align}
which is identical with Eq. \eqref{sfEOO} as should be.

\subsection{Spin-dependent Case}

The spin-orbit contributions to the orbital gradient and Hessian can be derived in precisely the same way as above. It is just that
the identities $z_{pq}^{(m)}=-z_{qp}^{(m)}$, $S^{(m)}_{pq}z^{(-m)}_{rs}=S^{(-m)*}_{qp}z^{(m)*}_{sr}$,
and $\sum_{m}S^{(m)}_{pq}z^{(-m)}_{rs}=\sum_{m}S^{(m)*}_{qp}z^{(-m)*}_{sr}$ should be used with care.
The following commutators
\begin{align}
a^{p\sigma}[a_{q\sigma}, \hat{T}_{rs}^{(m)}]&=\delta_{qr}\hat{T}^{(m)}_{ps},\\
[a^{p\sigma}, \hat{T}_{rs}^{(m)}]a_{q\sigma}&=-\delta_{ps}\hat{T}^{(m)}_{rq},\\
[\hat{E}_{pq}, \hat{T}^{(m)}_{rs}]&=\delta_{qr}\hat{T}^{(m)}_{ps}-\delta_{ps}\hat{T}^{(m)}_{rq}
\end{align}
can be used repeated to obtain the following expressions
\begin{align}
\hat{F}_{pq}^{\mathrm{so}}&=a^{p\sigma}[a_{q\sigma},\hat{h}^{\mathrm{so}}]=\sum_{m=-1}^{1}\hat{T}^{(m)}_{pz}Z^{(-m)}_{qz}
=-\sum_{m=-1}^{1}\hat{T}^{(m)}_{pz}Z^{(-m)}_{zq},\label{soKoopmans-op}\\
F_{pq}^{\mathrm{so}}&=\langle 0|\hat{F}_{pq}^{\mathrm{so}}|0\rangle
=-\sum_{m=-1}^{1}(\mathbf{S}^{(m)}\mathbf{Z}^{(-m)})_{pq},\quad S^{(m)}_{pq}=\langle 0|\hat{T}^{(m)}_{pq}|0\rangle,\label{soGenFock}\\
\hat{\tilde{F}}_{pq}^{\mathrm{so}}&=[a^{p\sigma},\hat{h}^{\mathrm{so}}]a_{q\sigma}=-(\hat{F}_{qp}^{\mathrm{so}})^\dag=-P_{pq}h.c.\hat{F}_{pq}^{\mathrm{so}}\\
&=-\sum_{m=-1}^{1}\hat{T}^{(m)}_{zq}Z^{(-m)}_{zp}=\sum_{m=-1}^{1}Z^{(-m)}_{pz}\hat{T}^{(m)}_{zq},\\
\tilde{F}_{pq}^{\mathrm{so}}&=\langle 0|\hat{\tilde{F}}_{pq}^{\mathrm{so}}|0\rangle=\sum_{m=-1}^{1}(\mathbf{Z}^{(-m)}\mathbf{S}^{(m)})_{pq}\\
&=\sum_{m=-1}^{1}(\mathbf{S}^{(m)}\mathbf{Z}^{(-m)})_{qp}^*=-P_{pq} F_{pq}^{\mathrm{so}*},\\
[\hat{E}_{pq},\hat{h}^{\mathrm{so}}]&=\hat{F}_{pq}^{\mathrm{so}}+\hat{\tilde{F}}_{pq}^{\mathrm{so}}=(1-P_{pq}h.c.)\hat{F}_{pq}^{\mathrm{so}},\\
&=\sum_{m=-1}^{1}(\hat{T}^{(m)}_{pz}Z^{(-m)}_{qz}-\hat{T}^{(m)}_{zq}Z^{(-m)}_{zp}),\\
[\hat{E}_{pq}^-,\hat{h}^{\mathrm{so}}]&=(1-\delta_{pq})(1-P_{pq})[\hat{E}_{pq},\hat{h}^{\mathrm{so}}]=(1-\delta_{pq})(1-P_{pq})(\hat{F}_{pq}^{\mathrm{so}}+h.c.)\label{soGRDop}\\
&=(1-\delta_{pq})(1-P_{pq})\sum_{m=-1}^{1}(\hat{T}^{(m)}_{pz}Z^{(-m)}_{qz}+\hat{T}^{(m)}_{zp}Z^{(-m)}_{zq}),\label{Epq-Hso}\\
G^{\mathrm{o,so}}_{pq}&=\langle 0| [E_{pq}^-, \hat{h}^{\mathrm{so}}]|0\rangle=G^{\mathrm{o,so}*}_{pq}=-G^{\mathrm{o,so}}_{qp}\\
&=2(1-\delta_{pq})(1-P_{pq})\Re (F_{pq}^{\mathrm{so}}),\label{soOrbGrd}\\
[\hat{E}_{pq}^-,\hat{F}_{rs}^{\mathrm{so}}]&=(1-\delta_{pq})(1-P_{pq})[\hat{E}_{pq},\hat{F}^{\mathrm{so}}_{rs}]\\
&=(1-\delta_{pq})(1-P_{pq})(\delta_{qr}\hat{F}_{ps}^{\mathrm{so}}-\sum_{m=-1}^{1}\hat{T}_{rq}^{(m)}Z^{(-m)}_{sp})\\
&=(1-\delta_{pq})(1-P_{pq})(\sum_{m=-1}^{1}\hat{T}_{rp}^{(m)}Z^{(-m)}_{sq}-\delta_{pr}\hat{F}_{qs}^{\mathrm{so}}).\label{soEFrs}
\end{align}
To calculate the spin-dependent orbital Hessian, we first invoke Eqs. \eqref{soGRDop} and \eqref{soEFrs} to obtain
\begin{align}
&[\hat{E}_{pq}^-,[\hat{E}_{rs}^-,\hat{h}^{\mathrm{so}}]]=(1-\delta_{pq})(1-\delta_{rs})(1-P_{rs})([\hat{E}_{pq}^-,\hat{F}_{rs}^{\mathrm{so}}]+h.c.)\\
&=2(1-\delta_{pq})(1-\delta_{rs})(1-P_{pq})(1-P_{rs})\nonumber\\
&\times \{-\frac{1}{2}\delta_{pr}(\hat{F}_{qs}^{\mathrm{so}}+h.c.)+\frac{1}{2}\sum_{m=-1}^{1}(\hat{T}^{(m)}_{pr}Z^{(-m)}_{qs}+h.c.) \},\label{soHessianop-nosym}
\end{align}
which can further be symmetrized to
\begin{align}
\hat{E}^{\mathrm{oo,so}}_{pq,rs}&=\frac{1}{2}(1+P_{pr}P_{qs})[\hat{E}_{pq}^-,[\hat{E}_{rs}^-,\hat{h}^{\mathrm{so}}]]\\
&=2(1-\delta_{pq})(1-\delta_{rs})(1-P_{pq})(1-P_{rs})\nonumber\\
&\times \{-\frac{1}{4}\delta_{pr}[(\hat{F}_{qs}^{\mathrm{so}}+\hat{F}_{sq}^{\mathrm{so}})+h.c.]+\frac{1}{2}\sum_{m=-1}^{1}(\hat{T}^{(m)}_{pr}Z^{(-m)}_{qs}+h.c.) \}.\label{soHessianop}
\end{align}
Since the above operator is manifestly Hermitian, its expectation is real-valued, viz.,
\begin{align}
E^{\mathrm{oo,so}}_{pq,rs}&=\langle 0|\hat{E}^{\mathrm{oo,so}}_{pq,rs}|0\rangle=-E^{\mathrm{oo,so}}_{qp,rs}=-E^{\mathrm{oo,so}}_{pq,sr}=E^{\mathrm{oo,so}}_{qp,sr}=E^{\mathrm{oo,so}*}_{pq,rs} \nonumber\\
&=2(1-\delta_{pq})(1-\delta_{rs})(1-P_{pq})(1-P_{rs})\nonumber\\
&\times \Re \{-\frac{1}{2}\delta_{pr} (F_{qs}^{\mathrm{so}}+F_{sq}^{\mathrm{so}})+\sum_{m=-1}^{1}S^{(m)}_{pr}Z^{(-m)}_{qs}\},\label{soEOO}
\end{align}
the diagonal term of which can be calculated as
\begin{align}
E^{\mathrm{oo,so}}_{pq,pq}&=\delta_{pr}\delta_{qs}E^{\mathrm{oo,so}}_{pq,rs}\\
&=-2(1-\delta_{pq})\Re\{(1+P_{pq})F_{qq}^{\mathrm{so}}+2S^{(m)}_{pq}Z^{(-m)}_{qp}\}\label{soDiagEOO}.
\end{align}
Again, Eq. \eqref{soDiagEOO} can only be obtained by acting $\delta_{pr}\delta_{qs}$ on the fully permuted expression of Eq. \eqref{soEOO}.

In summary, the full orbital gradient and Hessian read
\begin{align}
G^{\mathrm{o}}_{pq}&=G^{\mathrm{o,sf}}_{pq}+G^{\mathrm{o,so}}_{pq}=G^{\mathrm{o}*}_{pq}=-G^{\mathrm{o}}_{qp} \label{fullEOa}\\
&=2(1-\delta_{pq})(1-P_{pq})\Re F_{pq},\label{fullEO}\\
E^{\mathrm{oo}}_{pq,rs}&=E^{\mathrm{oo,sf}}_{pq,rs}+E^{\mathrm{oo,so}}_{pq,rs}=-E^{\mathrm{oo}}_{qp,rs}=-E^{\mathrm{oo}}_{pq,sr}=E^{\mathrm{oo}}_{qp,sr}=E^{\mathrm{oo}*}_{pq,rs}\\
&=2(1-\delta_{pq})(1-\delta_{rs})(1-P_{pq})(1-P_{rs})\nonumber\\
&\times \{\gamma_{pr}h^{\mathrm{sf}}_{qs}-\frac{1}{2}\delta_{pr}\Re (F_{qs}+F_{sq})+P^{xy}_{pr}J^{xy}_{qs}+2Q^{xy}_{pr}K^{xy}_{qs}\nonumber\\
&+\sum_{m=-1}^{1}S^{(m)}_{pr}Z^{(-m)}_{qs}\},\label{fullEOO}\\
E^{\mathrm{oo}}_{pq,pq}
&=2(1-\delta_{pq})\{(1+P_{pq})(-F_{qq}+\gamma_{pp}h^{\mathrm{sf}}_{qq}+P^{xy}_{pp}J^{xy}_{qq}+2Q^{xy}_{pp}K^{xy}_{qq})\nonumber\\
&-2(\gamma_{pq}h^{\mathrm{sf}}_{qp}+P^{xy}_{pq}J^{xy}_{qp}+2Q^{xy}_{pq}K^{xy}_{qp}+S^{(m)}_{pq}Z^{(-m)}_{qp})\},\label{fullDiagEOO}
\end{align}
where
\begin{align}
F_{pq}&=F_{pq}^{\mathrm{sf}}+F_{pq}^{\mathrm{so}}\\
&=D_{pz}h^{\mathrm{sf}}_{qz}+\Gamma_{pz,xy}g_{qz,xy}-\sum_{m=-1}^{1}(\mathbf{S}^{(m)}\mathbf{Z}^{(-m)})_{pq}.\label{fullFock}
\end{align}
\subsection{Further Factorization}
So far the manipulations are completely general. For the purpose of implementation,
further specifications of the various density matrices ($\mathbf{D}$, $\boldsymbol{\gamma}$,
$\boldsymbol{\Gamma}$, $\mathbf{P}$, $\mathbf{Q}$, and $\mathbf{S}^{(m)}$)
should be made. First of all, they
all vanish if any  of the indices is an unoccupied orbital. Moreover, the following relations also hold
\begin{align}
\hat{E}_{ip}|0\rangle&=\langle 0|\hat{E}_{pi}=2\delta_{ip},\quad |0\rangle=\sum_I|I\rangle C_I,\\
D_{pi}&=D_{ip}=2\delta_{ip},\\
\Gamma_{pq,is}&=\Gamma_{is,pq}=2\delta_{is}D_{pq}-\delta_{iq}D_{ps},\\
\Gamma_{pq,si}&=\Gamma_{si,pq}=2\delta_{is}D_{pq}-\delta_{ip}D_{sq},\\
\Gamma_{pu,iv}&=\Gamma_{iu,pv}=\Gamma_{uq,vi}=\Gamma_{ui,vs}=0,\\
%
S^{(m)}_{pi}&= S^{(m)}_{ip}=0, \label{Smatip}
\end{align}
in terms of which the non-Hermitian, generalized Fock matrix \eqref{fullFock} can be factorized as
\begin{align}
F_{iq}&=2f_{iq},\quad f_{pq}=f_{pq}^{\mathrm{c}}+f_{pq}^{\mathrm{a}},\label{FiqMat}\\
F_{tq}&=D_{tv}f^{\mathrm{c}}_{vq}+\Gamma_{tv,ef}g_{vq,ef}+F_{tq}^{\mathrm{so}},\label{FtqMat}\\
F_{aq}&=0,\label{FaqMat}\\
f_{pq}^{\mathrm{c}}&=h_{pq}^{\mathrm{sf}}+\sum_{k}(2J^{kk}_{pq} -K^{kk}_{pq})=f_{qp}^{\mathrm{c}},\\
f_{pq}^{\mathrm{a}}&=\sum_{tu}D_{tu}(J^{tu}_{pq} -\frac{1}{2}K^{tu}_{pq}),\quad \Re f_{pq}^{\mathrm{a}}=\Re f_{qp}^{\mathrm{a}},\\
F_{tq}^{\mathrm{so}}&=-\sum_{m=-1}^{1}S^{(m)}_{tv}Z^{(-m)}_{vq},\quad F_{iq}^{\mathrm{so}}=F_{aq}^{\mathrm{so}}=0.
\end{align}
Note in passing that, while $\Re F_{ij}$ is equal to $\Re F_{ji}$, $\Re F_{tu}$ is generally different from $\Re F_{ut}$.
The real-valued orbital gradient \eqref{fullEO} can then be factorized as
\begin{align}
G^{\mathrm{o}}_{ai} &=-2\Re F_{ia}=-4\Re f_{ia}=-G^{\mathrm{o}}_{ia},\label{GaiMat}\\
G^{\mathrm{o}}_{ti} &= 2\Re (F_{ti}-F_{it})=2\Re (F_{ti}-2f_{it})=-G^{\mathrm{o}}_{it},\\
G^{\mathrm{o}}_{at} &=-2\Re F_{ta}=-G^{\mathrm{o}}_{ta},\\
G^{\mathrm{o}}_{tu} &= 2\Re (F_{tu}-F_{ut})=-G^{\mathrm{o}}_{ut}.
\end{align}
To factorize the real-valued orbital Hessian \eqref{fullEOO}, the following identities
\begin{align}
P^{xy}_{pj}&=P^{xy}_{jp}=P_{xy}^{pj}=P_{xy}^{jp}=2\delta_{jp}\gamma_{xy}-\frac{1}{2}\delta_{jx}\gamma_{py}-\frac{1}{2}\delta_{jy}\gamma_{px},\\
Q^{xy}_{pj}&=Q^{yx}_{jp}=Q_{xy}^{pj}=Q_{yx}^{jp}=2\delta_{jy}\gamma_{px}-\frac{1}{2}\delta_{jp}\gamma_{xy}-\frac{1}{2}\delta_{jx}\gamma_{py}
\end{align}
and intermediates
\begin{align}
%
&P^{xy}_{tu}J^{xy}_{qs}+Q^{xy}_{tu}K^{xy}_{qs}=\gamma_{tu}(f^{\mathrm{c}}_{qs}-h^{\mathrm{sf}}_{qs})+P^{ef}_{tu}J^{ef}_{qs}+Q^{ef}_{tu}K^{ef}_{qs},\\
%
&P^{xy}_{pj}J^{xy}_{qs}+2Q^{xy}_{pj}K^{xy}_{qs}=2\delta_{pj}(f_{qs}-h^{\mathrm{sf}}_{qs})+2\delta_{pi}L^{ij}_{qs}+\gamma_{pv}L^{vj}_{qs},\\
&L^{xy}_{pq}=4K^{xy}_{pq}-K^{yx}_{pq}-J^{xy}_{pq}=L^{yx}_{qp}=L^{py}_{xq}=L^{xq}_{py}
\end{align}
can be invoked, such that
\begin{align}
E^{\mathrm{oo}}_{ai,bj}&=4\Re \{\delta_{ij}f_{ab}-\delta_{ab}f_{ij}+L^{ij}_{ab}\},\\
E^{\mathrm{oo}}_{ti,bj}&=4\Re \{\delta_{ij}f_{tb}-\frac{1}{4}\delta_{ij}F_{tb}+L^{ij}_{tb}\},\\
E^{\mathrm{oo}}_{at,bj}&=\Re\{-\delta_{ab}(F_{tj}+F_{jt})+2\gamma_{tv}L^{ab}_{vj}\},\\
E^{\mathrm{oo}}_{tu,bj}&=2(1-\delta_{tu})(1-P_{tu})\gamma_{uv}L^{tj}_{vb},\\
E^{\mathrm{oo}}_{ti,uj}&=2\Re \{2\delta_{ij}f_{tu}-2\delta_{tu}f_{ij}-\frac{1}{2}\delta_{ij}(F_{tu}+F_{ut})\nonumber\\
&+(\delta_{tw}-\gamma_{tw})L^{ij}_{wu}+L^{ij}_{tw}(\delta_{wu}-\gamma_{wu})+\gamma_{tu}f_{ij}^{\mathrm{c}}+P_{tu}^{ef}J_{ij}^{ef}+2Q_{tu}^{ef}K^{ef}_{ij}\nonumber\\
&+\sum_{m=-1}^{1}S^{(m)}_{tu}Z^{(-m)}_{ij}\},\\
E^{\mathrm{oo}}_{at,uj}&=2\Re \{\delta_{tu}f_{aj}+\gamma_{tv}L^{vu}_{aj}-(\gamma_{tu}f^{\mathrm{c}}_{aj}+P^{ef}_{tu}J^{ef}_{aj}+2Q^{ef}_{tu}K^{ef}_{aj})\nonumber\\
&-\sum_{m=-1}^{1}S^{(m)}_{tu}Z^{(-m)}_{aj}\},\\
E^{\mathrm{oo}}_{tw,uj}&=2(1-\delta_{tw})(1-P_{tw})\Re \{-\frac{1}{2}\delta_{tu}(F_{wj}+F_{jw})-\gamma_{tv}L^{vu}_{wj}\nonumber\\
&+\gamma_{tu}f^{\mathrm{c}}_{wj}+P^{ef}_{tu}J^{ef}_{wj}+2Q^{ef}_{tu}K^{ef}_{wj}+\sum_{m=-1}^{1}S^{(m)}_{tu}Z^{(-m)}_{wj}\},\\
E^{\mathrm{oo}}_{at,bu}&=2\Re \{-\frac{1}{2}\delta_{ab} (F_{tu}+F_{ut})+\gamma_{tu} f_{ab}^{\mathrm{c}} +P^{ef}_{tu}J^{ef}_{ab}+2Q^{ef}_{tu}K^{ef}_{ab}\nonumber\\
&+\sum_{m=-1}^{1}S^{(m)}_{tu}Z^{(-m)}_{ab}\},\\
E^{\mathrm{oo}}_{tw,bu}&=2(1-\delta_{tw})(1-P_{tw})\{-\frac{1}{2}\delta_{wu}F_{tb}+\gamma_{wu}f^{\mathrm{c}}_{tb}+P^{ef}_{wu}J^{ef}_{tb}+2Q^{ef}_{wu}K^{ef}_{tb}\nonumber\\
&+\sum_{m=-1}^{1}S^{(m)}_{wu}Z^{(-m)}_{tb}\},\\
E^{\mathrm{oo}}_{tu,vw}&=2(1-\delta_{tu})(1-\delta_{vw})(1-P_{tu})(1-P_{vw})\nonumber\\
&\times \Re\{-\frac{1}{2}\delta_{tv} (F_{uw}+F_{wu}) +\gamma_{tv}f^{\mathrm{c}}_{uw} +P^{ef}_{tv}J^{ef}_{uw}+2Q^{ef}_{tv}K^{ef}_{uw}\nonumber\\
&+\sum_{m=-1}^{1}S^{(m)}_{tv}Z^{(-m)}_{uw}\}.
\end{align}
The diagonal elements \eqref{fullDiagEOO} of the orbital Hessian read
\begin{align}
E^{\mathrm{oo}}_{ai,ai}&=4\Re \{f_{aa}-f_{ii}+L^{ii}_{aa} \},\\
E^{\mathrm{oo}}_{ti,ti}&=2\Re \{2f_{tt}-2f_{ii} - F_{tt}+2 (\delta_{tw}-\gamma_{tw}) L^{ii}_{wt} \nonumber\\
           &+\gamma_{tt}f_{ii}^{\mathrm{c}}+P^{ef}_{tt}J^{ef}_{ii}
          +2 Q^{ef}_{tt}K^{ef}_{ii}\},\\
E^{\mathrm{oo}}_{at,at}&=2\Re \{- F_{tt} + \gamma_{tt} f_{aa}^{\mathrm{c}} + 2P^{ef}_{tt}J^{ef}_{aa}+ 2Q^{ef}_{tt}K^{ef}_{aa}\},\\
E^{\mathrm{oo}}_{tu,tu} &=2(1-\delta_{tu})\Re \{(1+P_{tu})(-F_{uu}+\gamma_{tt}f^{\mathrm{c}}_{uu} +P^{ef}_{tt}J^{ef}_{uu}+2Q^{ef}_{tt}K^{ef}_{uu})\nonumber\\
&-2(\gamma_{tu}f^{\mathrm{c}}_{ut}+P^{ef}_{tu}J^{ef}_{ut}+2Q^{ef}_{tu}K^{ef}_{ut}+S^{(m)}_{tu}Z^{(-m)}_{ut})\}.
\end{align}

Several remarks are in order:
\begin{enumerate}[(1)]
\item The $F^{\mathrm{so}}_{tq}$-containing terms have been absorbed into
the generalized Fock matrix $F_{tq}$ \eqref{FtqMat},
such that only the $S^{(m)}Z^{(-m)}$ terms of $G_{pq}^{\mathrm{o,so}}$ \eqref{soOrbGrd} and
$E^{\mathrm{oo,so}}_{pq,rs}$ \eqref{soEOO} appear explicitly in the total $G_{pq}^{\mathrm{o}}$ and $E^{\mathrm{oo}}_{pq,rs}$, respectively.
Rather unexpectedly, the SOC has no affect on the rotations between doubly occupied and virtual orbitals (see Eq. \eqref{GaiMat} and Appendix \ref{PseudoSingles} for further discussions).
\item As far as molecular symmetry is concerned for the evaluation of $S^{(m)}_{tv}Z^{(-m)}_{vq}$, we have $\Gamma_{t}=\Gamma_{q}$ but $\Gamma_{v}$
is related to the irrep $\Gamma_{R_{l}}$ of the rotations $R_{l}$ ($l = x, y, z$, i.e.,
the cartesian components $(H_{\mathrm{so}}^{l})_{vq}$ in $Z^{(-m)}_{vq}$), that is, $\Gamma_{v}=\Gamma_{R_{l}}\otimes \Gamma_{q}$
for binary point groups. Moreover, both double binary groups ($D_{2h}^*$ and its subgroups) and time reversal symmetries can be
incorporated\cite{SOiCI} for the evaluation of the magnetic density matrices $\mathbf{S}^{(m)}$.
\item It is well known\cite{Quaternion1983,Quaternion1999} that, for
$C_1^*$ and $C_i^*$,  the CI matrix has a quaternion structure as long as the many-electron basis is time reversal adapted, such that it can be block-diagonalized by a quaternion unitary transformation,
thereby reducing the order of the CI matrix from $2N$ to $N$.
However, there practically is no gain in the partial diagonalization step (which scales as $mN^2$ for $m$ roots) by working with quaternion algebra,
because four complex multiplications are needed to perform one quaternion multiplication.
Moreover, the implementation of quaternion algebra is much less efficient than that of complex algebra.
\end{enumerate}

Finally, Eq. \eqref{SymmE} can be factorized as
\begin{align}
E^{(0)}=h^\mathrm{sf}_{ii}+f^{\mathrm{c}}_{ii}+\gamma_{tu}f^{\mathrm{c}}_{ut}
+\frac{1}{2}{P}^{ef}_{tu}J^{ef}_{ut}-\sum_{m=-1}^{1}\mathbf{S}^{(m)}_{tu}\mathbf{Z}^{(-m)}_{ut}.\label{FinalE}
\end{align}
\section{Pseudo-singles}\label{PseudoSingles}
\setcounter{equation}{0}
\renewcommand{\theequation}{B\arabic{equation}}

As indicated by Eq. \eqref{Smatip}, the matrix elements $S^{(m)}_{pq}=\langle 0|\hat{T}^{(m)}_{pq}|0\rangle$ can be nonzero only
when both $p$ and $q$ are active orbitals, such that the SOC does \emph{not} introduce any rotations between doubly occupied and unoccupied,
real-valued orbitals.
Noticing that the so-called pseudo-single excitations $\{\hat{E}_{at}\hat{E}_{ti}\}$
also generate CSFs of the same spin as those by $\{E_{ai}\}$, it is attempted to augment $\{\kappa_{ai}\}$ with the following term
\begin{equation}
\sum_{ai} \kappa_{ai}\sum_t (\hat{E}_{at}\hat{E}_{ti}-\hat{E}_{it}\hat{E}_{ta})=\sum_{ai}\kappa_{ai}\tilde{E}_{ai}^-.\label{PseudoS}
\end{equation}
However, its contribution to the corresponding orbital gradient is still null, as can be seen as follows
\begin{align}
&\langle 0|[\tilde{E}_{ai}^-, \hat{h}^{\mathrm{so}}]|0\rangle
=-\sum_{m=-1}^{1} \sum_{rt} \{Z^{(-m)}_{ra}\langle 0|\hat{T}^{(m)}_{rt}E_{ti}|0\rangle+Z^{(-m)}_{ar}\langle 0|E_{it}\hat{T}^{(m)}_{tr}|0\rangle\}\\
&=-\sum_{m=-1}^{1} \sum_{rt} \{Z^{(-m)}_{ra}\langle 0|\hat{T}^{(m)}_{ri}-\delta_{ir}\hat{T}^{(m)}_{tt}|0\rangle
+Z^{(-m)}_{ar}\langle 0|\hat{T}^{(m)}_{ir}-\delta_{ir}\hat{T}^{(m)}_{tt}|0\rangle\}\\
&=\sum_{m=-1}^{1} \sum_{t}(Z^{(-m)}_{ia}+Z^{(-m)}_{ai})\langle 0|\hat{T}^{(m)}_{tt}|0\rangle =0.
\end{align}
In contrast, the term \eqref{PseudoS} may have nonvanishing contribution to the spin-free orbital gradient, e.g.,
\begin{eqnarray}
G^{\mathrm{o,sf}}_{ai}&\leftarrow&\langle 0|[\tilde{E}_{ai}^-, \hat{h}^{\mathrm{sf}}]|0\rangle\\
&=&- \sum_{rt} \{h^{\mathrm{sf}}_{ra}\langle 0|\hat{E}_{rt}\hat{E}_{ti}|0\rangle+h_{ar}^{\mathrm{sf}}\langle 0|\hat{E}_{it}\hat{E}_{tr}|0\rangle\}\\
&=&- \sum_{rt} \{h^{\mathrm{sf}}_{ra}\langle 0|\hat{E}_{ri}-\delta_{ir}\hat{E}_{tt}|0\rangle+h_{ar}^{\mathrm{sf}}\langle 0|\hat{E}_{ir}-\delta_{ir}\hat{E}_{tt}|0\rangle\}\\
&=&2h^{\mathrm{sf}}_{ai}\sum_{t}(n_t-2),\quad n_t=\langle 0|\hat{E}_{tt}|0\rangle.
\end{eqnarray}
However, it is in either case illegal to invoke terms like Eq. \eqref{PseudoS} for orbital optimization, simply
because the unitary transformation of orbitals dictates that only one-body excitation
operators should appear in $\hat{\kappa}$ \eqref{kappadef}, viz.,
\begin{eqnarray}
\tilde{a}^{p\sigma}=e^{-\hat{\kappa}} a^{p\sigma} e^{\hat{\kappa}}= a^{r\sigma} (e^{-\boldsymbol{\kappa}})_{rp}=a^{r\sigma} U_{rp}.
\end{eqnarray}
It is obvious that the operator $\hat{E}_{at}\hat{E}_{ti}$ does not lead to the above form:
the (particle) rank of $[a^{p\sigma}, \hat{E}_{at}\hat{E}_{ti}]$ is $3/2$ instead of the desired $1/2$ for $\tilde{a}^{p\sigma}$.

\section*{Supporting Information}
The Supporting Information
No.

\bibliography{iCI}

\newpage
For TOC only

\includegraphics[width=\textwidth]{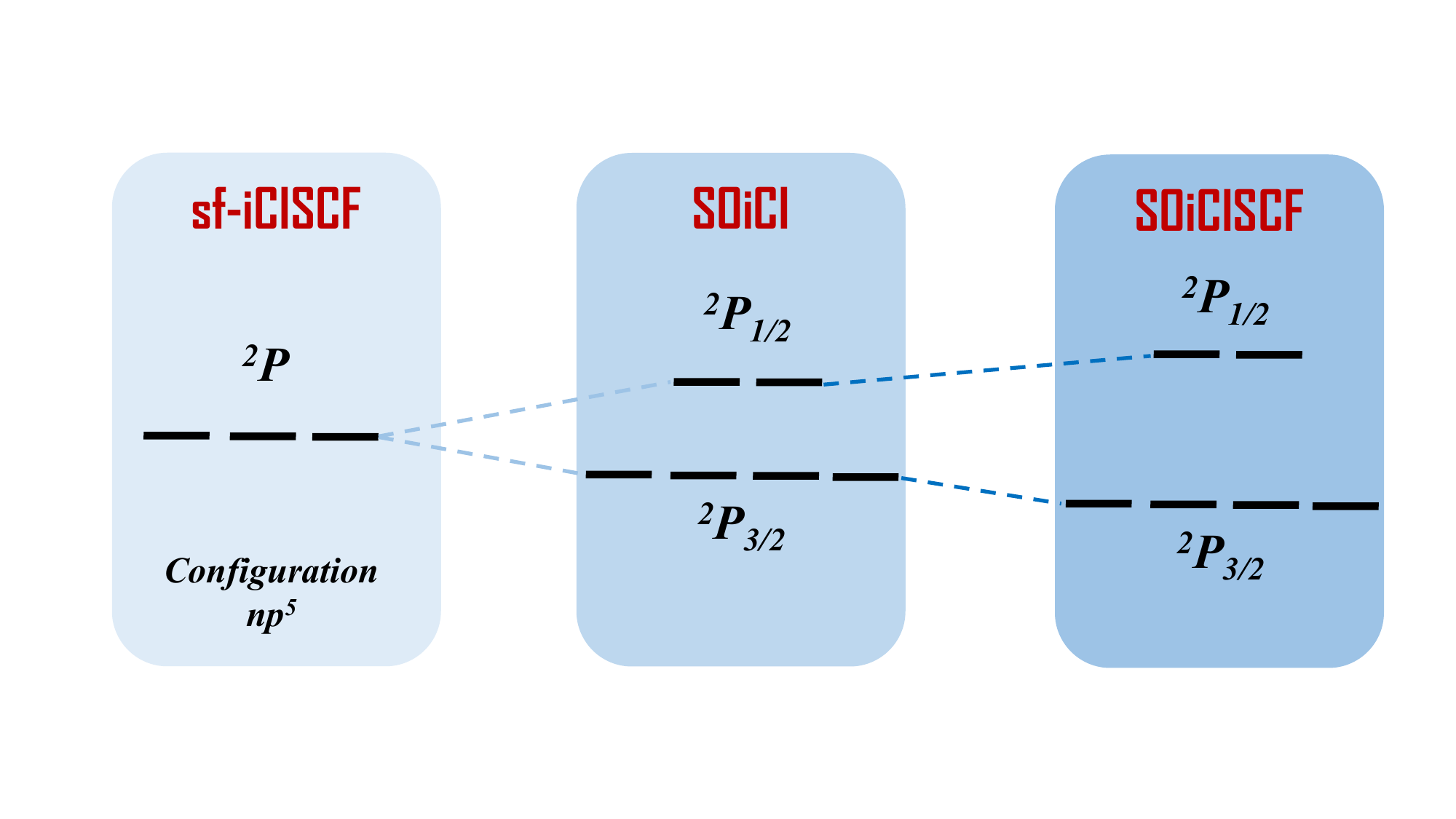}

\end{document}


\renewcommand{\thetable}{S\arabic{table}}
\renewcommand{\thefigure}{S\arabic{figure}}

\begin{center}
{\Large Supporting Information}	
\end{center}
\newpage

\section{spin-orbit splittings (SOSs) of selected $\textit{p}$-block elements computed by SOiCI(2) and SOiCISCF(2).}

\begin{table}[!htp]
	\footnotesize
	\scriptsize
	\centering
	\caption{The SOSs (in cm$^{-1})$ of At with the vtz-$m$p$^1_2$ basis sets. The results are provided for two cases: when the 12-13$p$ orbitals are deleted in SOiCISCF calculations by VSD, and when they are included as virtual MOs. The percentage errors w.r.t. the mmf-X2C-FS-CCSD result are given in parentheses.}
	\begin{threeparttable}
		\centering
		\begin{tabular}{c|cccc}\toprule
			\multirow{2}{*}{$^2P_{1/2}$$-$$^{2}P_{3/2}$}& \multicolumn{2}{c}{without 12-13$p$} & \multicolumn{2}{c}{with 12-13$p$} \\
			& \multicolumn{1}{c}{SOiCI(2)} & SOiCISCF(2)
			& \multicolumn{1}{c}{SOiCI(2)} & SOiCISCF(2)\\\toprule
			CAS(5, 3)    &20272 (-13.35\%) &20272 (-13.35\%) &  20272 (-13.35\%)  & 20272 (-13.35\%)  \\ 
			CAS(5, 6)    &21487 (-8.15\%)  &41041 (75.43\%)  &  21909 (-6.35\%)   & 112211 (379.66\%)  \\
			CAS(11, 9)   &21571 (-7.79\%)  &20937 (-10.50\%) &  23063 (-1.42\%)   & 24091 (2.98\%)    \\ 
			CAS(11, 12)  &21950 (-6.17\%)  &22381 (-4.33\%)  &  24010 (2.63\%)    & 28295 (20.95\%)   \\ 
			CAS(11, 15)  &22055 (-5.72\%)  &22432 (-4.11\%)  &  23476 (0.35\%)    & 26499 (13.27\%)   \\ 
			CAS(11, 18)  &22077 (-5.63\%)  &22509 (-3.78\%)  &  25336 (8.30\%)    & 28989 (23.91\%)   \\ 
			CAS(11, 21)  &22532 (-3.69\%)  &22541 (-3.65\%)  &  26130 (11.69\%) &   not converge\\                   
			\\                                                                                          
			CAS(13, 14)  &22059 (-5.71\%)  &22479 (-3.91\%)  &  24194 (3.42\%)   &   28040 (19.86\%)\\                 
			CAS(15, 16)  &21929 (-6.26\%)  &22625 (-3.29\%)  &  23763 (1.58\%)   &   27981 (19.61\%)\\                 
			CAS(25, 26)  &20833 (-10.95\%) &23000 (-1.69\%)  &  21186 (-9.44\%)  &   29425 (25.78\%)\\                 
			CAS(25, 33)  &20684 (-11.58\%) &22735 (-2.82\%)  &  21059 (-9.98\%)  &   31353 (34.02\%)\\                 
			mmf-X2C-FS-CCSD & \multicolumn{4}{c}{23394} \\\midrule
		\end{tabular}
	\end{threeparttable}\label{At-no-delete}	
\end{table}

\begin{table}[!htp]
	\footnotesize
	\centering
	\caption{The SOSs (in cm$^{-1})$ of selected icosagens. The percentage errors w.r.t. experimental values are given in parentheses.}
	\begin{threeparttable}
		\centering
		\begin{tabular}{c|llllll}\toprule
			\multirow{2}{*}{$^2P_{3/2}$$-$$^{2}P_{1/2}$}&\multicolumn{2}{c}{Ga}  & \multicolumn{2}{c}{In} & \multicolumn{2}{c}{Tl} \\
			& \multicolumn{1}{c}{SOiCI(2)} & SOiCISCF(2)
			& \multicolumn{1}{c}{SOiCI(2)} & SOiCISCF(2)
			& \multicolumn{1}{c}{SOiCI(2)} & SOiCISCF(2)\\\toprule
			CAS(1,3)     &712 (-13.88\%) &712 (-13.88\%) &1907 (-13.82\%) &1907 (-13.82\%) &6463 (-17.06\%) &6463 (-17.06\%) \\
			CAS(7,9)    &794 (-3.93\%) &789 (-4.54\%) &2024 (-8.53\%) &2144 (-3.11\%) &6867 (-11.88\%) &7264 (-6.79\%) \\
			CAS(7,12)   &782 (-5.40\%) &783 (-5.24\%) &2231 (0.82\%) &2157 (-2.49\%) &6761 (-13.23\%) &6824 (-12.43\%) \\
			CAS(9,14)   &726 (-12.15\%) &721 (-12.72\%) &1874 (-15.30\%) &1984 (-10.35\%) &6553 (-15.91\%) &7215 (-7.42\%) \\
			CAS(21,26)   &788 (-4.65\%) &824 (-0.31\%) &2176 (-1.67\%) &2413 (9.07\%) &6094 (-21.80\%) &7054 (-9.48\%) \\
			CAS(21,33)   &800 (-3.14\%) &837 (1.32\%) &2223 (0.45\%) &2472 (11.70\%) &6077 (-22.02\%) &6992 (-10.27\%) \\
			CAS(21,full)   &780 (-5.58\%) &780 (-5.57\%) &2170 (-1.93\%) &2167 (-2.06\%) &7040 (-9.66\%) &6967 (-10.59\%) \\
			Exp.  &\multicolumn{2}{c}{826}	&	\multicolumn{2}{c}{2213}	&	\multicolumn{2}{c}{7793}\\\midrule
		\end{tabular}
	\end{threeparttable}\label{Ga2Tl}
\end{table}

\begin{table}[!htp]
	\footnotesize
	\centering
	\caption{The SOSs (in cm$^{-1})$ of selected carbon group elements. The percentage errors w.r.t. experimental values are given in parentheses.}
	\begin{threeparttable}
		\centering
		\begin{tabular}{c|llllll}\toprule
			&\multicolumn{2}{c}{Ge}  & \multicolumn{2}{c}{Sn} & \multicolumn{2}{c}{Pb} \\
			& \multicolumn{1}{c}{SOiCI(2)}& SOiCISCF(2)
			& \multicolumn{1}{c}{SOiCI(2)} & SOiCISCF(2)
			& \multicolumn{1}{c}{SOiCI(2)} & SOiCISCF(2)\\\toprule
			&\multicolumn{6}{c}{$^3P_{1}$$-$$^{3}P_{0}$}\\
			CAS(2,3)      &470 (-15.56\%) &473 (-15.10\%) &1367 (-19.21\%) &1418 (-16.19\%) &6199 (-20.72\%) &8241 (5.39\%) \\
			CAS(8,9)      &533 (-4.34\%) &529 (-5.04\%) &1391 (-17.80\%) &1505 (-11.05\%) &6351 (-18.78\%) &7224 (-7.61\%) \\
			CAS(8,12)     &520 (-6.72\%) &523 (-6.20\%) &1538 (-9.08\%) &1510 (-10.76\%) &6684 (-14.52\%) &6670 (-14.70\%) \\
			CAS(10,14)    &525 (-5.81\%) &520 (-6.64\%) &1440 (-14.90\%) &1552 (-8.28\%) &7352 (-5.97\%) &7585 (-2.99\%) \\
			CAS(22,26)    &500 (-10.28\%) &522 (-6.36\%) &1521 (-10.08\%) &1533 (-9.40\%) &6549 (-16.25\%) &7399 (-5.37\%) \\
			CAS(22,33)    &501 (-10.15\%) &522 (-6.21\%) &1469 (-13.15\%) &1472 (-13.01\%) &6695 (-14.38\%) &7314 (-6.46\%) \\
			CAS(22,full)    &489 (-12.28\%) &490 (-11.98\%) &1410 (-16.68\%) &1439 (-14.93\%) &6129 (-21.61\%) &7114 (-9.02\%) \\
			Exp.  &\multicolumn{2}{c}{557}	&	\multicolumn{2}{c}{1692}	&	\multicolumn{2}{c}{7819}\\
			\\
			&\multicolumn{6}{c}{$^3P_{2}$$-$$^{3}P_{0}$} \\
			CAS(2,3)      &1250 (-11.37\%) &1256 (-10.94\%) &3084 (-10.03\%) &3177 (-7.33\%) &9773 (-8.24\%) &12097 (13.58\%) \\
			CAS(8,9)      &1398 (-0.88\%) &1388 (-1.53\%) &3129 (-8.71\%) &3333 (-2.77\%) &10299 (-3.30\%) &11631 (9.21\%) \\
			CAS(8,12)     &1367 (-3.08\%) &1373 (-2.59\%) &3391 (-1.06\%) &3342 (-2.49\%) &10576 (-0.70\%) &11021 (3.49\%) \\
			CAS(10,14)    &1345 (-4.63\%) &1335 (-5.35\%) &3112 (-9.22\%) &3307 (-3.53\%) &11073 (3.97\%) &11521 (8.18\%) \\
			CAS(22,26)    &1371 (-2.79\%) &1417 (0.48\%) &3308 (-3.49\%) &3354 (-2.16\%) &10461 (-1.78\%) &11656 (9.45\%) \\
			CAS(22,33)    &1377 (-2.32\%) &1424 (1.00\%) &3356 (-2.10\%) &3394 (-0.99\%) &10519 (-1.23\%) &11791 (10.71\%) \\
			CAS(22,full)   &1358 (-3.72\%) &1358 (-3.67\%) &3316 (-3.26\%) &3334 (-2.74\%) &11217 (5.32\%) &11265 (5.77\%) \\
			Exp.  &\multicolumn{2}{c}{1410}	&	\multicolumn{2}{c}{3428}	&	\multicolumn{2}{c}{10650}\\\midrule
		\end{tabular}
	\end{threeparttable}\label{Ge2Pb}
\end{table}

\begin{table}[!htp]
	\footnotesize
	\centering
	\caption{The SOSs (in cm$^{-1})$ of selected elements in group 15. The percentage errors w.r.t experimental values are given in parentheses.}
	\begin{threeparttable}
		\centering
		\begin{tabular}{c|llllll}\toprule
			&\multicolumn{2}{c}{As}  & \multicolumn{2}{c}{Sb} & \multicolumn{2}{c}{Bi} \\
			& \multicolumn{1}{c}{SOiCI(2)} & SOiCISCF(2)
			& \multicolumn{1}{c}{SOiCI(2)} & SOiCISCF(2)
			& \multicolumn{1}{c}{SOiCI(2)} & SOiCISCF(2)\\\toprule
			&\multicolumn{6}{c}{$^2D_{5/2}$$-$$^{2}D_{3/2}$}\\
			CAS(3,3)     &188 (-41.77\%) &188 (-41.59\%) &984 (-26.71\%) &984 (-26.71\%) &4321 (7.52\%) &4322 (7.56\%) \\
			CAS(9,9)     &229 (-28.91\%) &226 (-29.80\%) &1006 (-25.01\%) &1096 (-18.36\%) &4579 (13.96\%) &4444 (10.60\%) \\
			CAS(9,12)    &220 (-31.65\%) &222 (-31.02\%) &1115 (-16.91\%) &1207 (-10.02\%) &4805 (19.57\%) &4799 (19.41\%) \\
			CAS(11,14)   &406 (25.98\%) &400 (24.17\%) &1501 (11.85\%) &1615 (20.37\%) &5064 (26.01\%) &5205 (29.53\%) \\
			CAS(23,26)   &304 (-5.74\%) &311 (-3.56\%) &1855 (38.21\%) &1994 (48.62\%) &4582 (14.03\%) &4912 (22.23\%) \\
			CAS(23,33)   &284 (-11.71\%) &292 (-9.38\%) &1697 (26.43\%) &1225 (-8.71\%) &4165 (3.65\%) &4473 (11.31\%)  \\
			CAS(23,full)  &284 (-11.79\%) &284 (-11.82\%) &1237 (-7.81\%) &1238 (-7.74\%) &4371 (8.78\%) &4314 (7.35\%) \\	
			Exp.  &\multicolumn{2}{c}{322}	&	\multicolumn{2}{c}{1342}	&	\multicolumn{2}{c}{4019}\\
			\\
			&\multicolumn{6}{c}{$^2P_{3/2}$$-$$^{2}P_{1/2}$} \\
			CAS(3,3)     &250 (-45.92\%) &250 (-45.75\%) &1366 (-33.98\%) &1366 (-33.98\%) &8333 (-27.56\%) &8324 (-27.65\%) \\
			CAS(9,9)     &305 (-34.00\%) &301 (-34.75\%) &1403 (-32.17\%) &1533 (-25.92\%) &9178 (-20.22\%) &8837 (-23.18\%) \\
			CAS(9,12)    &292 (-36.71\%) &295 (-36.02\%) &1658 (-19.87\%) &1807 (-12.63\%) &10338 (-10.13\%) &10419 (-9.43\%) \\
			CAS(11,14)   &444 (-3.86\%) &437 (-5.34\%) &1890 (-8.66\%) &2048 (-0.98\%) &10310 (-10.38\%) &10507 (-8.66\%) \\
			CAS(23,26)   &365 (-20.97\%) &366 (-20.78\%) &1803 (-12.87\%) &1827 (-11.69\%) &9268 (-19.44\%) &10507 (-8.67\%) \\
			CAS(23,33)   &356 (-22.75\%) &357 (-22.68\%) &1717 (-17.01\%) &1663 (-19.62\%) &9217 (-19.88\%) &10478 (-8.92\%) \\
			CAS(23,full)  &381 (-17.49\%) &382 (-17.17\%) &1750 (-15.40\%) &1757 (-15.08\%) &10490 (-8.81\%) &10545 (-8.34\%) \\
			Exp.  &\multicolumn{2}{c}{461}	&	\multicolumn{2}{c}{2069}	&	\multicolumn{2}{c}{11504}\\\midrule
		\end{tabular}
	\end{threeparttable}\label{As2Bi}
\end{table}

\begin{table}[!htp]
	\footnotesize
	\centering
	\caption{The SOSs (in cm$^{-1})$ of selected chalcogens. The percentage errors w.r.t. experimental values are given in parentheses.}
	\begin{threeparttable}
		\centering
		\begin{tabular}{c|llllll}\toprule
			&\multicolumn{2}{c}{Se}  & \multicolumn{2}{c}{Te} & \multicolumn{2}{c}{Po} \\
			& \multicolumn{1}{c}{SOiCI(2)} & SOiCISCF(2)
			& \multicolumn{1}{c}{SOiCI(2)} & SOiCISCF(2)
			& \multicolumn{1}{c}{SOiCI(2)} & SOiCISCF(2)\\\toprule
			&\multicolumn{6}{c}{$^3P_{1}$$-$$^{3}P_{2}$}\\
			CAS(4, 3)	  &1830 (-8.04\%) &1842 (-7.41\%) &4224 (-11.09\%) &4347 (-8.50\%) &14316 (-14.95\%) &16964 (0.78\%) \\
			CAS(10, 9)  &2028 (1.95\%) &2076 (4.34\%) &4275 (-10.01\%) &4747 (-0.08\%) &15185 (-9.78\%) &18125 (7.68\%) \\
			CAS(10, 12)	&1987 (-0.12\%) &1986 (-0.17\%) &4303 (-9.42\%) &4551 (-4.21\%) &15177 (-9.83\%) &15815 (-6.04\%) \\
			CAS(12, 14)	&2007 (0.88\%) &2005 (0.77\%) &4328 (-8.90\%) &4576 (-3.69\%) &15638 (-7.09\%) &16166 (-3.96\%) \\
			CAS(24, 26)	&1921 (-3.47\%) &1975 (-0.71\%) &4454 (-6.24\%) &4762 (0.25\%) &14817 (-11.97\%) &16345 (-2.89\%) \\
			CAS(24, 33)	&1916 (-3.69\%) &1971 (-0.94\%) &4379 (-7.82\%) &4497 (-5.34\%) &14795 (-12.10\%) &16456 (-2.23\%) \\
			CAS(24, full)&1860 (-6.50\%) &1865 (-6.28\%) &4164 (-12.35\%) &4187 (-11.86\%) &14858 (-11.73\%) &15725 (-6.58\%) \\
			Exp.  &\multicolumn{2}{c}{1990}	&	\multicolumn{2}{c}{4751}	&	\multicolumn{2}{c}{16832}\\
			\\
			&\multicolumn{6}{c}{$^3P_{0}$$-$$^{3}P_{2}$} \\
			CAS(4, 3)	   &2453 (-3.44\%) &2468 (-2.85\%) &4795 (1.87\%) &4895 (4.01\%) &9081 (20.84\%) &9512 (26.58\%) \\
			CAS(10, 9)   &2689 (5.82\%) &2743 (7.96\%) &4833 (2.68\%) &5211 (10.71\%) &9217 (22.66\%) &9653 (28.45\%) \\
			CAS(10, 12)  &2616 (2.95\%) &2615 (2.92\%) &4859 (3.23\%) &5063 (7.57\%) &8193 (9.03\%)  &8368 (11.36\%) \\
			CAS(12, 14)  &2564 (0.90\%) &2562 (0.83\%) &4601 (-2.24\%) &4734 (0.59\%) &7440 (-0.99\%) &7468 (-0.62\%) \\
			CAS(24, 26)  &2585 (1.73\%) &2650 (4.31\%) &4926 (4.67\%) &5151 (9.44\%) &8470 (12.71\%) &8854 (17.82\%) \\
			CAS(24, 33)  &2593 (2.05\%) &2659 (4.64\%) &4825 (2.52\%) &5028 (6.84\%) &8708 (15.87\%) &9151 (21.77\%) \\
			CAS(24, full) &2564 (0.89\%) &2552 (0.43\%) &4876 (3.61\%) &4811 (2.22\%) &13547 (80.27\%) &8049 (7.11\%) \\
			Exp.  &\multicolumn{2}{c}{2541}	&	\multicolumn{2}{c}{4707}	&	\multicolumn{2}{c}{7515}\\\midrule
		\end{tabular}
	\end{threeparttable}\label{Se2Po}
\end{table}